\documentclass[%
aip,
amsmath,amssymb,
reprint,%
]{revtex4-1}

\usepackage{graphicx}
\usepackage{dcolumn}
\usepackage{bm}

\usepackage[utf8]{inputenc}
\usepackage[T1]{fontenc}
\usepackage{mathptmx}
\usepackage{etoolbox}
\usepackage[normalem]{ulem}

\makeatletter
\def\@email#1#2{%
	\endgroup
	\patchcmd{\titleblock@produce}
	{\frontmatter@RRAPformat}
	{\frontmatter@RRAPformat{\produce@RRAP{*#1\href{mailto:#2}{#2}}}\frontmatter@RRAPformat}
	{}{}
}%
\makeatother
\begin{document}
	
	\preprint{AIP/123-QED}
	
	\title{Efficient Random Phase Approximation for Diradicals}
	\author{Reza G. Shirazi}
	
	\author{Vladimir V. Rybkin}
	
	\author{Michael Marthaler}

\author{Dmitry S. Golubev}
	\affiliation{$^1$HQS Quantum Simulations GmbH, Rintheimer Str. 23, 76131 Karlsruhe, Germany}


\begin{abstract}
We apply the analytically solvable model of two electrons in two orbitals to diradical molecules, characterized by two unpaired electrons.
The effect of the doubly occupied and empty orbitals is taken into account by means of random phase approximation (RPA).
We show that in the static limit the direct RPA leads to the renormalization of the parameters of the two-orbital model.
We test our model by comparing its predictions for the singlet-triplet splitting with the results from multi-reference CASSCF and NEVPT2 simulations for a set of ten molecules.
We find that, for the whole set, the average relative difference between the singlet-triplet gaps predicted by the RPA-corrected two-orbital model and by NEVPT2 is about 40\%.
For the five molecules with the smallest singlet-triplet splitting the accuracy is better than 20\%.
\end{abstract}

\maketitle

\section{Introduction}
The model of two electrons localized in two orbitals is relatively simple and
can be solved analytically (see, for example, Refs. \cite{Scully,Malrieu,Broer}). 
This model plays an important role in quantum chemistry as it
provides a simple description for the wide class of diradical molecules having two unpaired electrons \cite{Abe,Stuyver}. 
It is equally important for solid state physics where it is used to describe magnetic materials \cite{Broer}
and double quantum dots \cite{Loss,Burkard,Hu}, which are currently
actively studied due to their potential applications in quantum computing and quantum optics.  

It has been realized long ago that the two-orbital model in its simplest form
is insufficient to describe the properties of diradicals on the quantitative level.
To improve the accuracy, one should consider the interactions with other orbitals
and electrons.
It has been also shown that such interactions can often be incorporated 
into the two-orbital model by replacing the original model
parameters with the renormalized ones. The renormalized parameters can be calculated, for example,
by means of the perturbation theory \cite{Calzado,Malrieu}.
In this paper we build on these ideas and propose to obtain 
the renormalized parameters of the two-orbital model by means of the random phase approximation (RPA).
This approximation goes further than simple perturbation theory and, at the same time,
it remains computationally cheap. Our approach can also be
used to improve the modelling of semiconducting double quantum dots, for which the 
two-orbital model can be used \cite{Loss,Burkard,Hu}.
 
The RPA approximation has been first proposed 
by Bohm and Pines \cite{Bohm1,Bohm2,Bohm3} to describe strongly interacting electrons in solid state physics. 
Later it was generalized for inhomogeneous systems like atoms and molecules \cite{Ball,Dunning}.  
Nowadays, the RPA approximation became one of the main tools of calculating the
correlation energy of molecules \cite{Furche1,Furche2,Furche3,Hesselmann,Rinke,Paris}.
Here we will use the static screening limit of the RPA approximation. In this limit,
the effect of the doubly occupied orbitals reduces to the screening of Coluomb
interaction between the two singly occupied orbitals of the diradical.
It is well known that in the homogeneous electron gas such screening leads to the
replacement of the bare Coulomb potential $e^2/r$, where $e$ is the electrom charge
and $r$ is the distance, by the Yukawa potential of the form $(e^2/r)\exp[-r/\lambda_{\rm TF}]$, where
$\lambda_{\rm TF}$ is the Thomas-Fermi screening length. However, since the charge
distribution in a molecule is very inhomogeneous, this simple formula
cannot be used, and one should describe the screening
effect in terms of the  two-particle Coulomb integrals.
Below we will work out the corresponding expressions.
The importance of the RPA corrections for diradical molecules has been earlier emphasized in Ref. \cite{Yang}. 
However Ref. \cite{Yang} is focused on numerical methods, while our 
goal here is to provide a simple, yet accurate, analytical model for a diradical.

The singlet-triplet gap, $\Delta E_{\rm ST}$, is one of the important factors defining the chemistry of diradicals.
Here we use the convention that $\Delta E_{\rm ST}$ equals to the energy difference between the singlet and the triplet states of the diradical,
which means $\Delta E_{\rm ST}<0$ corresponds to the singlet ground state and $\Delta E_{\rm ST}>0$ implies the triplet ground state. In general,
the ground state of a diradical cannot be easily found with the black-box single reference
methods unless one exploits unrestricted mean-field approaches breaking the spin symmetry and assuming different orbitals for different spins \cite{Neese}. 
These methods are hard to converge to the right state and they are prone to heavy spin contamination, 
sacrificing correct wave function properties for the sake of decent energetics \cite{Yang_ocigoacenes}. 

The usual robust way 
of finding the gap $\Delta E_{\rm ST}$, and the spin symmetry of the ground state, is based on the second order perturbation theory
methods such as the complete active space second order perturbation theory  (CASPT2) and the $N$-electron
valence state second-order perturbation theory (NEVPT2). These methods are usually applied on top of the orbitals optimized
with the complete active space self consistent field (CASSCF) method 
with active spaces containing  non-bonding orbitals and  $\pi$-orbitals \cite{Borden}. 
The non-bonding orbitals are usually the degenerate singly occupied radical orbitals, and they we will be the main focus
of our further discussion. The goal of the present work is to improve the two-orbital model for a diradical
such that it predicts reasonably accurate values for the singlet-triplet gap $\Delta E_{\rm ST}$. 
We show that the static limit of the direct RPA approximation achieves this goal.

Over the years, a lot of experimental and theoretical research has been devoted to diradicals and double quantum dots. 
For example, in a recent experiment \cite{Dressler} a number of polycyclic aromatic hydrocarbon diradicaloids 
have been synthesized, and the authors have demonstrated the ability to tune the singlet-triplet gap in a controllable manner
by varying the geometry of the molecules. 
Moreover, in semi-conductor double quantum dots one can tune all system parameters, including the
gap $\Delta E_{\rm ST}$, very precisely applying gate voltages \cite{Petta}.
Such tunablity of the gap provides a good testing ground for analytically tracktable models.
On the theory side, it has been shown that the properties of diradicals are well modeled numerically  
if one uses the triplet state, well approximated by a single Slater determinant, as a reference, 
and creates the correlated singlet state by applying a spin flip operation to the triplet \cite{Krylov,Krylov_TMM,Wang,Lee}. 
This was an additional motivation for our study. Indeed, since the triplet state 
of a diradical is weakly correlated, the correlations in the singlet 
state are unlikely to be very strong, and a linear combination of only two properly chosen
independent Slater determinants should be sufficient to describe them. 
In this case, a two-orbital model with renormalized parameters should capture such correlations. 

The paper is organized as follows: 
In Sec. \ref{model} we define the Hamiltonian of the molecule in a very general form and 
split it into the Hamiltonian of the two diradical orbitals, the "environment"
Hamiltonian including all other orbitals, both doubly occupied and empty,
and the interaction terms between these two sub-systems.
In Sec. \ref{Sec_HF} we perform the averaging over the environment orbitals on the level
of the Hartree-Fock approximation and provide the analytical solution of the
resulting two-orbital model of the diradical.
In Sec. \ref{Sec_RPA} we introduce the direct RPA approximation, consider its static
limit and show how it modifies the parameters of the two-orbital model.
In Sec. \ref{Sec_results} we compare the predictions of the model
with the simulations based on CASSCF and NEVPT2 methods and
in Sec. \ref{Sec_conlcusion} we summarize our results. 
Details of the calculations are presented in the Appendices.

%
%
\section{Theory}

\subsection{System Hamiltonian}
\label{model}

We start from the very general form of the Hamiltonian for a molecule, 
\begin{eqnarray}
H = \sum_\sigma\sum_{pq=1}^N t_{pq} c^\dagger_{p\sigma} c_{q\sigma} 
+  \sum_{\sigma\sigma'} \sum_{pqrs=1}^N \frac{h_{psqr}}{2} c^\dagger_{p\sigma} c^\dagger_{q\sigma'} c_{r\sigma'} c_{s\sigma}. 
\label{H}
\end{eqnarray}
Here $c_{q\sigma}^\dagger, c_{q\sigma}$ are the fermionic creation and annihillation operators of an electron with spin $\sigma$ on the orbital $q$,
$N$ is the total number of orbitals (ideally $N$ tends to infinity),
\begin{eqnarray}
t_{pq} = \int d^3{\bm r}\; \psi_p({\bm r})\left[ -\frac{\hbar^2\nabla^2}{2m^*} + U({\bm r}) \right]\psi_q({\bm r}),
\label{t0pq}
\end{eqnarray}
are the matrix elements of the free electron Hamiltonian between the wave functions of the orbitals $\psi_p({\bm r})$ and $\psi_q({\bm r})$,
$m^*$ is the electron mass, $U({\bm r})$ is the potential induced by the atomic nuclei, and
\begin{eqnarray}
h_{psqr} = \int d^3{\bm r} d^3{\bm r}' \; \psi_p({\bm r})\psi_s({\bm r})\frac{e^2}{|{\bm r}-{\bm r}'|}\psi_q({\bm r}')\psi_r({\bm r}')
\label{h_int}
\end{eqnarray}
are the so-called "chemical" Coulomb integrals, i.e.  $h_{psqr} = (ps|qr)$. 

Since here we are considering diradicals, 
we separate the two orbitals hosting the two unpaired electrons and number them as the orbitals 1 and 2. 
All remaining orbitals, both doubly occupied and empty, are numbered as $3,4,5,\dots, N$. 
These orbitals form the environment for the diradical orbitals.
After such separation, the Hamiltonian (\ref{H}) takes the form
\begin{eqnarray}
H = H_{12} + H_{\rm env} + V,
\end{eqnarray}
where $H_{12}$ is the Hamiltonian of the two selected orbitals,
\begin{eqnarray}
H_{12} = \sum_\sigma\sum_{pq=1}^2 t_{pq} c^\dagger_{p\sigma} c_{q\sigma} 
+  \sum_{\sigma\sigma'} \sum_{pqrs=1}^2 \frac{h_{psqr}}{2} c^\dagger_{p\sigma} c^\dagger_{q\sigma'} c_{r\sigma'} c_{s\sigma}, 
\label{H12}
\end{eqnarray}
$H_{\rm env}$ is the Hamiltonian of the environment, 
\begin{eqnarray}
H_{\rm env} = \sum_\sigma\sum_{pq=3}^N t_{pq} c^\dagger_{p\sigma} c_{q\sigma} 
+  \sum_{\sigma\sigma'} \sum_{pqrs=3}^N \frac{h_{psqr}}{2} c^\dagger_{p\sigma} c^\dagger_{q\sigma'} c_{r\sigma'} c_{s\sigma}, 
\label{Henv}
\end{eqnarray}
and $V$ is the interaction term between the diradical orbitals and the environment. $V$ includes all terms of the Hamiltonian (\ref{H}), 
in which some of the indices take the values 1 or 2 and the remaining indices refer to the environment orbitals. 
The interaction term can be split in several parts. Employing the symmteris of the integrals $t_{pq}$ 
and $h_{psqr}$, we express $V$ as
\begin{eqnarray}
V = V_t + V_{\rm C} + V_{\rm ex} + V_{\rm pair} + V_1 + V_3,
\end{eqnarray}
where
\begin{eqnarray}
V_t = \sum_\sigma\sum_{p=1}^2 \sum_{r=3}^N t_{pr} ( c^\dagger_{p\sigma} c_{r\sigma} + c^\dagger_{r\sigma} c_{p\sigma} )
\label{V_t}
\end{eqnarray}
describes the hopping between the two selected orbitals and the environment,
\begin{eqnarray}
V_{\rm C} = \sum_{p,q=1}^2 \sum_{r,s=3}^N h_{pqrs} 
\left( c^\dagger_{p\uparrow} c_{q\uparrow} + c^\dagger_{p\downarrow} c_{q\downarrow} \right) 
\left( c^\dagger_{r\uparrow} c_{s\uparrow} + c^\dagger_{r\downarrow} c_{s\downarrow} \right) 
\label{V_C}
\end{eqnarray}
is the Coulomb interaction between the orbitals 1,2 and the environment,
\begin{eqnarray}
V_{\rm ex} &=& -  \sum_{\sigma\sigma'} \sum_{p,q=1}^2 \sum_{r,s=3}^N 
\frac{h_{psrq}}{2} \big( c^\dagger_{p\sigma} c_{q\sigma'} c^\dagger_{r\sigma'} c_{s\sigma} 
\nonumber\\ &&
+\, c^\dagger_{s\sigma} c_{r\sigma'} c^\dagger_{q\sigma'} c_{p\sigma} \big)
\label{V_ex}
\end{eqnarray}
is the exchange interaction between the two sub-systems,
\begin{eqnarray}
V_{\rm pair} &=&  \sum_{\sigma\sigma'} \sum_{p,q=1}^2 \sum_{r,s=3}^N 
\frac{h_{psqr}}{2} \big(c^\dagger_{p\sigma} c^\dagger_{q\sigma'} c_{r\sigma'} c_{s\sigma} 
\nonumber\\ &&
+\, c^\dagger_{s\sigma} c^\dagger_{r\sigma'} c_{q\sigma'} c_{p\sigma} \big)
\label{V_pair}
\end{eqnarray}
describes the hopping of pairs of electrons between the orbitals 1,2 and the environmnet,
and in $V_1$ and $V_3$ we collect the terms having either one or three fermionic operators corresponding to the orbitals 1 and 2,
\begin{eqnarray}
V_1 = \sum_{\sigma\sigma'} \sum_{p=1}^2 \sum_{krs=3}^N 
h_{pskr} \left(  c^\dagger_{p\sigma} c^\dagger_{k\sigma'} c_{r\sigma'} c_{s\sigma} +  c^\dagger_{s\sigma} c^\dagger_{r\sigma'} c_{k\sigma'} c_{p\sigma} \right),
\label{V1}
\\
V_3 = \sum_{\sigma\sigma'} \sum_{pql=1}^2 \sum_{k=3}^N 
h_{klpq} \left( c^\dagger_{k\sigma} c^\dagger_{p\sigma'} c_{q\sigma'} c_{l\sigma} + c^\dagger_{l\sigma} c^\dagger_{q\sigma'} c_{p\sigma'} c_{k\sigma}  \right).
\label{V3}
\end{eqnarray}

\subsection{Simple two-orbital model}
\label{Sec_HF}

In this section we perform the averaging over the environment orbitals with the aid of Wick's theorem.
This corresponds to Hartree-Fock approximation for these orbitals.
In contrast, we treat the singly occupied orbitals 1 and 2 exactly. 

The wave functions of the orbitals 1 and 2 are assumed to be orthogonal, $\langle \psi_p | \psi_q \rangle = \delta_{pq}$, 
for $p,q = 1,2$. We also choose the basis of the orbitals in such a way that the single particle density matrix 
of the environment is diagonal,
\begin{eqnarray}
\rho_{rs} = \langle c^\dagger_{r\uparrow} c_{s\uparrow} + c^\dagger_{r\downarrow} c_{s\downarrow} \rangle = 2n_r \delta_{rs}, \;\;\; r,s\geq 3.
\label{rho}
\end{eqnarray}
Here $n_r = \langle c^\dagger_{r\uparrow} c_{r\uparrow} \rangle = \langle c^\dagger_{r\downarrow} c_{r\downarrow} \rangle$ 
are the occupation numbers of the environment orbitals, which are assumed to be spin-degenerate.
In the Hartree-Fock approximation one finds $n_r=1$ for the doubly occupied orbitals and $n_r=0$ for the empty ones.
After the averaging over the enviroment operators the interaction terms $V_t,V_1,V_3$ and $V_{\rm pair}$ vanish
and the Hamiltonian (\ref{H}) takes the form
\begin{eqnarray}
H_{\rm av} = E_{\rm env}^{\rm HF} + H_{12},
\label{H0}
\end{eqnarray}
where   
\begin{eqnarray}
E_{\rm env}^{\rm HF} = 2\sum_{q=3}^N t_{qq} n_q + \sum_{pq=3}^N ( 2h_{ppqq} - h_{pqqp} )n_pn_q
\label{Eenv}
\end{eqnarray}
is the Hartree-Fock energy of the environment,
\begin{eqnarray}
H_{12} = \sum_\sigma\sum_{pq=1}^2 t'_{pq} c^\dagger_{p\sigma} c_{q\sigma}
+  \sum_{\sigma\sigma'} \sum_{pqrs=1}^2 \frac{h_{psqr}}{2} c^\dagger_{p\sigma} c^\dagger_{q\sigma'} c_{r\sigma'} c_{s\sigma}
\label{H012}
\end{eqnarray}
is the two-orbital Hamiltonian, and
\begin{eqnarray}
t'_{pq} = t_{pq} + \sum_{k=3}^N (2h_{pqkk}- h_{pkkq})n_k
\label{tilde_t}
\end{eqnarray}
are the hopping ampltidues between the two singly occupied orbitals modified by the interaction with the environment.

The two-orbital Hamiltonian (\ref{H012}) is exactly solvable, see e.g. Refs. \cite{Malrieu,Broer,Nakano,He} etc. 
Here we briefly summarize the solution epxressing the result in a compact form applicable
to any choice of the orbital wave functions $\psi_1,\psi_2$, for any hopping matrix elements $t'_{pq}$
and for any set of Coulmb integrals $h_{pqrs}$.

We define the parameters of the model as
\begin{eqnarray}
&& U_1 = \frac{h_{1111}}{2}, \;\; U_2 = \frac{h_{2222}}{2}, \;\; J_{12} = h_{1122}, \;\; K_{12} = h_{1212},
\nonumber\\ &&
t_1 = t'_{12} + h_{1112},\;\; {\rm and} \;\; t_2 = t'_{12} + h_{1222}.
\label{UJK} 
\end{eqnarray}
They have the following physical meaning: $U_1$ and $U_2$ are the charging energies of the orbitals,
$J_{12}$ is the direct Coulomb interaction between the orbitals,  $K_{12}$ is the exchange integral,
$t_1$ is the hopping amplitude between the orbitals 1 and 2 if before or after the hopping the orbital 1
was doubly occupied, and $t_2$ is the similar amplitude corresponding to the case when the orbital 2 was doubly occupied.
We also introduce the energies of the orbitals 1 and 2 shifted by the Hartree-Fock interaction with the environment, $\varepsilon_1 = t'_{11}$
and $\varepsilon_2 = t'_{22}$.
The full Hilbert space for the two-orbital Hamiltonian (\ref{H012}) consists of 6 states:
\begin{eqnarray}
&& \Psi_1 = | \uparrow\downarrow, 0\rangle = c_{1\uparrow}^\dagger c_{1\downarrow}^\dagger |0\rangle, \;\;\; 
\Psi_2 = | 0 , \uparrow\downarrow \rangle = c_{2\uparrow}^\dagger c_{2\downarrow}^\dagger |0\rangle, \;\;\;
\nonumber\\ &&
\Psi_3 = | \uparrow, \downarrow\rangle = c_{1\uparrow}^\dagger c_{2\downarrow}^\dagger |0\rangle, \;\;\; 
\Psi_4 = | \downarrow, \uparrow \rangle = c_{1\downarrow}^\dagger c_{2\uparrow}^\dagger |0\rangle, \;\;\;
\nonumber\\ &&
\Psi_5 = | \uparrow, \uparrow \rangle = c_{1\uparrow}^\dagger c_{2\uparrow}^\dagger |0\rangle, \;\;\; 
\Psi_6 = | \downarrow, \downarrow \rangle = c_{1\downarrow}^\dagger c_{2\downarrow}^\dagger |0\rangle.
\end{eqnarray} 
In this basis it takes the form of the $6\times 6$ matrix with the block-diagonal structure
\begin{eqnarray}
H_{12} = \left( \begin{array}{cc} H_{12}^{(0)} & 0 \\ 0 & H_{12}^{\pm 1} \end{array} \right).
\end{eqnarray} 
The $4\times 4$ block $H_{12}^{(0)}$ connects the states $\Psi_1,\Psi_2,\Psi_3$ and $\Psi_4$ having zero $z$-component of the total spin
(here $E_{12}=\varepsilon_1 + \varepsilon_2$), 
\begin{eqnarray}
H_{12}^{(0)}=\left(\begin{array}{cccc} 
2 \varepsilon_1+ 2U_1 & K_{12} & t_1 & -t_1   \\
K_{12} & 2\varepsilon_2+2U_2 & t_2 & -t_2    \\
t_1 & t_2 & E_{12} + J_{12} & -K_{12}   \\
-t_1 & -t_2 & -K_{12} & E_{12} + J_{12}    
\end{array}\right),
\nonumber\\
\label{H20}
\end{eqnarray}
and the diagonal $2\times 2$ block $H_{12}^{\pm 1}$ describes the states $\Psi_5$ and $\Psi_6$ with $z$-components of the total spin equal to +1 and to -1,
\begin{eqnarray}
H_{\rm av}^{\pm 1}=\left(\begin{array}{cc} 
E_{12} + J_{12} - K_{12} & 0    \\
0 & E_{12} + J_{12} - K_{12}    
\end{array}\right).
\label{H2pm}
\end{eqnarray} 
Diagonalizing the block $H_{12}^{(0)}$ (\ref{H20}), we find its' 4 eigen-energies.
One of them coincides with the two eigenenergies of the diagonal matrix $H_{\rm av}^{\pm 1}$, and together they form the triplet state. 
The total energy of the molecule in the triplet state, i.e. the corresponding eigenvalue of the Hamiltonian (\ref{H0}), is 
\begin{eqnarray}
E^{\rm tr} = E_{\rm env}^{\rm HF} + \varepsilon_1 + \varepsilon_2 + J_{12} - K_{12}, 
\label{Etr}
\end{eqnarray} 
and the three triplet wave functions are
\begin{eqnarray}
| \uparrow, \uparrow \rangle, \;\;\; (| \uparrow, \downarrow \rangle+| \downarrow, \uparrow \rangle)/\sqrt{2} , \;\;\; | \downarrow, \downarrow \rangle.
\end{eqnarray}
To obtain the three remaining eigenenergies of $H_{12}^{(0)}$, which correspond to the three separate singlet states, 
one has to solve the third order polynomial equation
\begin{eqnarray}
&&(\varepsilon_1 + \varepsilon_2 + J_{12} + K_{12} - E)\big[ (2\varepsilon_1 + 2U_1 - E) 
\nonumber\\  && \times\,
(2\varepsilon_2 + 2U_2 - E) - K_{12}^2 \big] 
- 2t_1^2(2\varepsilon_2 + 2U_2 - E) 
\nonumber\\ &&
-\, 2t_2^2 (2\varepsilon_1 + 2U_1 - E) + 4t_1t_2 K_{12} = 0.
\label{polynom}
\end{eqnarray}
This can be done analytically with the aid of Cardano's formula. 
To make the notations more compact, 
we introduce the following paramters:
the splitting between the energies of the orbitals modified by Coulomb interaction, 
\begin{eqnarray}
\delta\varepsilon = \varepsilon_1 + U_1 - \varepsilon_2 - U_2,
\end{eqnarray}
the exchange integral between the symmetric and the antisymmetric combinations of the orbital wave functions 
$\psi^*_{1,2} = (\psi_1 \pm \psi_2)/\sqrt{2}$, 
\begin{eqnarray}
K^*_{12} = \frac{U_1 + U_2 - J_{12}}{2},
\end{eqnarray}
the two combinations of the exchange intergals and the hopping amplitudes, 
\begin{eqnarray}
K_0 = 2K^*_{12} -K_{12}, \;\;
K' = K_0 - \frac{2t_1t_2 K_{12} + (t_1^2-t_2^2)\delta\varepsilon}{t_1^2+t_2^2},
\end{eqnarray}
the hopping amplitude $t_{0} = \sqrt{t_1^2+t_2^2}$
and, finally, the effective level splitting modified by the exchange interaction,
\begin{eqnarray}
\Delta_0 = \sqrt{ \frac{K_0^2}{3} + K_{12}^2 + 2t_{0}^2 + \delta\varepsilon^2 }.
\end{eqnarray}
In terms of these parameters, the three singlet eigenenergies of the full Hamiltonian (\ref{H0}) are expressed as (here $j=1,2,3$)
\begin{eqnarray}
&& E_j^{\rm sg} = E_{\rm env}^{\rm HF} + \varepsilon_1 + U_1 + \varepsilon_2 + U_2 - \frac{K_0}{3} - \frac{2\Delta_0}{\sqrt{3}}\times
\nonumber\\ &&
\cos\bigg[ \frac{2\pi j}{3}
+ \frac{1}{3}\arccos\left( \frac{4K_0^3}{\sqrt{27}\Delta_0^3} - \frac{\sqrt{3}K_0}{\Delta_0} 
+ \frac{\sqrt{27} t_{0}^2 K'}{\Delta_0^3} \right) 
\bigg].
\nonumber\\ 
\label{Esg}
\end{eqnarray}
Here we used the representation of Cardano's formula in terms of trigonometric functions.
The lowest singlet energy is $E_3^{\rm sg}$, therefore the singlet-triplet splitting is
given by the expression
\begin{eqnarray}
&& \Delta E_{\rm ST} = E_3^{\rm sg} - E^{\rm tr} = 
2K_{12} + \frac{2K_0}{3} - \frac{2\Delta_0}{\sqrt{3}}\times\,
\nonumber\\ &&
\cos\left[ \frac{1}{3}\arccos\left( \frac{4K_0^3}{\sqrt{27}\Delta_0^3} - \frac{\sqrt{3}K_0}{\Delta_0} 
+ \frac{\sqrt{27} t_{0}^2 K'}{\Delta_0^3} \right)  \right].
\label{dEST}
\end{eqnarray}

Using the relation
\begin{eqnarray}
	\cos\left[\frac{\arccos (4x^3-3x)}{3}\right] = 
	\left\{ \begin{array}{cc} \frac{\sqrt{3(1-x^2)} - x}{2}, &  x<1/2, \\ x, & x>1/2, \end{array}\right.
	\label{identity}
\end{eqnarray} 
we can discuss several limiting cases for the singlet-triplet splitting.
First, if the hopping between the orbitals is suppressed, i.e. if $t_1=t_2=0$,
one can put $x=K_0/\sqrt{3}\Delta_0$ and the identity (\ref{identity}) leads to \cite{Hay}
\begin{eqnarray}
\Delta E_{\rm ST} &=&  
2K^*_{12}+K_{12} - \sqrt{ K_{12}^2 + \delta\varepsilon^2 },\;{\rm if}\;  K^*_{12} < F(K_{12},\delta\varepsilon), 
\nonumber\\
\Delta E_{\rm ST} &=& 2K_{12}, \;{\rm if}\; K^*_{12} > F(K_{12},\delta\varepsilon).
\label{dE_t0}
\end{eqnarray} 
Here we have defined the function $F(K_{12},\delta\varepsilon)=( K_{12} + \sqrt{ K_{12}^2 + \delta\varepsilon^2 } )/2$.
This expression applies, for example, to symmetric diatomic molecules in the basis of "gerade" and "ungerade" orbtial wave functions.
Second, for a symmetric system with $t_1=t_2=t$, $\varepsilon_1=\varepsilon_2=\varepsilon$ and $U_1=U_2=U$
one can put $x=(K^*_{12} + K_{12})/\sqrt{3}\Delta_0$, and Eqs. (\ref{dEST},\ref{identity}) give  \cite{Hay,Calzado}
\begin{eqnarray}
\Delta E_{\rm ST} &=& 2K_{12} + K^*_{12} - \sqrt{ K_{12}^{*2} + 4t^2  }, \;{\rm if}\; K_{12}  > F(K^*_{12},2t), 
\nonumber\\
\Delta E_{\rm ST} &=& 2K^*_{12}, \;{\rm if}\; K_{12} < F(K^*_{12},2t).
\label{dE_t}
\end{eqnarray}
We emphasize the symmetry between the expressions (\ref{dE_t0}) and (\ref{dE_t}). Namely, 
interchanging the parameters $K^*_{12} \leftrightarrow K_{12}$ and $\delta\varepsilon\leftrightarrow 2t$,
one obtains Eq. (\ref{dE_t0}) from Eq. (\ref{dE_t}) and vice versa.
The third simple limit is a rotationally invariant system 
in which the properties $K^*_{12} = K_{12}$ and $t_1=t_2=t$ hold. In this case, one finds $K_0=K_{12}$, $K' = 0$
and, by setting $x=K_{12}/\sqrt{3}\Delta_0$, from Eqs. (\ref{dEST},\ref{identity}) one obtains
\begin{eqnarray}
\Delta E_{\rm ST} = 3K_{12} - \sqrt{K_{12}^2 + 4t^2 + \delta\varepsilon^2}.
\end{eqnarray}
Finally, in the absence of Coulomb interaction one finds $K_{12}=K_0=K'=0$, $U_1=U_2=0$ and $t_1=t_2= t'_{12}$. In this case,
the energies (\ref{Etr},\ref{Esg}) become $E^{\rm tr} = E_2^{\rm sg} = E_{\rm env}^{\rm HF} +\varepsilon_1+\varepsilon_2$, 
$E_{1,3}^{\rm sg}=E_{\rm env}^{\rm HF} +\varepsilon_1+\varepsilon_2 \pm \sqrt{4t_{12}^{\prime\, 2} + (\varepsilon_1 - \varepsilon_2)^2}$ and the singlet-triplet
splitting (\ref{dEST}) equals to 
\begin{eqnarray}
\Delta E_{\rm ST} = - \sqrt{4 t_{12}^{\prime\, 2} + (\varepsilon_1 - \varepsilon_2)^2}.
\end{eqnarray}
It is always negative, which implies the singlet ground state with two electrons occupying the
lowest energy orbital.

\subsection{Direct RPA approximation}
\label{Sec_RPA}

In the previous section the effect of the environment on the diradical orbitals 1 and 2
has been taken into account by means of the Hartree-Fock approximation, which modifies only
the non-interacting part of the two-orbital Hamiltonian (\ref{H012}) via the relation (\ref{tilde_t}).
The corresponding corrections contain the Coulomb integrals with pairs of coinciding indexes of the type $h_{ii kk}$ and $h_{ikik}$.
In this section we introduce the direct RPA approximation, which takes into account further corrections coming from
the Coulomb interaction term (\ref{V_C}). These corrections contain the Coulomb integrals
of the type $h_{pq m\alpha}$, where the indexes $p$ and $q$ take the values 1 or 2, while the indexes $m,\alpha$ refer to
the environment orbitals, $m,\alpha\geq 3$. Here and below the Greek indices ($\alpha,\beta$) refer to the doubly occupied orbitals
and the Latin ones ($m,n$) -- to the empty orbitals. 

The RPA approximation replaces every excitation of an electron from a doubly occupied orbital $\alpha$
to an empty orbital $m$ by an excitation of an oscillator with the frequency
\begin{eqnarray}
\hbar\omega_{m\alpha} =  t'_{mm} - t'_{\alpha\alpha} - h_{mm\alpha \alpha } + h_{m\alpha m\alpha},
\label{omega}
\end{eqnarray}
which corresponds to the difference between the Hartree-Fock energies of the corresponding excited state and the ground state of the molecule.
Here we have introduced the Hartree-Fock operator for the environment orbitals, which extends Eq. (\ref{tilde_t}) to the environment indexes $r,s\geq 3$,
\begin{eqnarray}
t'_{rs} = t_{rs} + \sum_{k=3}^N (2h_{rskk}-h_{rksk})n_k, \;\;\; r,s\geq 3.
\label{tpq}
\end{eqnarray} 
This operator is diagonal, $t'_{rs} = t'_{rr}\delta_{rs}$, if Hartree-Fock approximation 
or CASSCF with two orbitals in the active space (CASSCF(2,2)) was used to find the environment orbitals.
The RPA approximation is expected to work well as long as the second and the third excited states of the oscillators are weakly
populated, which formally requires sufficiently weak interaction. 
Technically, the direct RPA approximation ammounts to
the following replacement of the opertors in the Hamiltonians (\ref{Henv},\ref{V_C}):
\begin{eqnarray}
c_{m\uparrow}^\dagger c_{\alpha\uparrow} + c_{m\downarrow}^\dagger c_{\alpha\downarrow} &\rightarrow & \sqrt{2} a_{m\alpha}^\dagger ,
\nonumber\\
c_{\alpha\uparrow}^\dagger c_{m\uparrow} + c_{\alpha\downarrow}^\dagger c_{m\downarrow} &\rightarrow & \sqrt{2} a_{m\alpha}.
\label{rep}
\end{eqnarray}
Here $a^\dagger_{m\alpha}$ is the ladder operator of the oscillator corresponding to the single electron excitation $\alpha\to m$.
Thus, the direct RPA accounts for the Coulomb interaction term $V_{\rm C}$ (\ref{V_C}) between the diradical orbitals
and the environment. However, it ignores 
the exchange term $V_{\rm ex}$ (\ref{V_ex}), the pair interaction term $V_{\rm pair}$ (\ref{V_pair}) and the term $V_3$ (\ref{V3}).
These terms can be omitted if the diradical orbitals 1 and 2 are separated from the envrionment
orbitals by sufficiently large energy gap.
Next, the term $V_1$ can be treated within the Hartree-Fock approximation. Applying Wick's theorem to it 
and replacing the pairs of operators 
$c^\dagger_{k\sigma'} c_{r\sigma'}$ or $c^\dagger_{k\sigma'} c_{s\sigma}$  by their averages, \textit{i.e.}
by $n_k\delta_{kr}$ and $n_k\delta_{ks}\delta_{\sigma\sigma'}$ respectively,
we notice that $V_1$ (\ref{V1}) can be combined with $V_t$ (\ref{V_t}) into
the modified hopping term $V'_t =V_t+V_1$. It has the same form as $V_t$, but
the hopping amplitude $t_{pr}$ in it is replaced by the Hartree-Fock corrected amplitude $t'_{pq}$ (\ref{tpq}).
Here we assume that the matrix (\ref{tpq}) is diagonal, which is the case, if the Hartree-Fock or CASSCF(2,2)
was used to obtain the Hamiltonian (\ref{H}) and the integrals (\ref{t0pq},\ref{h_int}).
Therefore, below we put $V_t+V_1 = 0$.

After these approximations, the original Hamiltonian (\ref{H}) is replaced by the following one
\begin{eqnarray}
H = H_{12} + H_{\rm env} + V_{\rm RPA}.
\label{H11}
\end{eqnarray}
Here,  $H_{12}$ is the two-orbital Hamiltonian (\ref{H012}), 
\begin{eqnarray}
H_{\rm env} &=& E_{\rm env}^{\rm HF} 
+\sum_{m,\alpha} \left( \hbar\omega_{m\alpha} a^\dagger_{m\alpha}a_{m\alpha} - h_{m\alpha m\alpha} \right)
\nonumber\\ &&
+\, \sum_{m\alpha,n\beta} h_{m\alpha n\beta} (a^\dagger_{m\alpha} + a_{m\alpha}) (a^\dagger_{n\beta} + a_{n\beta})
\label{HRPA}
\end{eqnarray} 
is the Hamiltonian of the environment orbitals (\ref{Henv}) in which the replacement (\ref{rep}) has been made, and
\begin{eqnarray}
V_{\rm RPA} = \sqrt{2} \sum_{\sigma=\uparrow,\downarrow}\sum_{m\alpha}\sum_{p,q=1}^2 h_{pqm\alpha} (a^\dagger_{m\alpha} + a_{m\alpha}) c^\dagger_{p\sigma} c_{q\sigma}
\label{VRPA}
\end{eqnarray}
is the Coulomb interaction term $V_{\rm C}$ (\ref{V_C}) within the RPA approximation. 
The summation in Eqs. (\ref{HRPA}) and (\ref{VRPA})
runs over all possible pairs of the occupied  and empty  orbitals $\alpha,m$ and $\beta,n$. 
The Hamiltonian of the environment contains the counterterm $\propto h_{m\alpha m\alpha}$ given by the last term in the first line of Eq. (\ref{HRPA}).
It can be derived either by inegration over the coupling constant \cite{Furche2}
or by comparing the lowest three eigenenergies of the Hamiltonian $H_{\rm env}$ in the two orbital limit, i.e. for $\alpha=1$ and $m=2$, 
with the three singlet eigenenergies of the two-orbital Hamiltonian (\ref{Esg}).
This comparison also helps to establish the
formal criteria of validity of the direct RPA approximation, namely,
\begin{eqnarray}
h_{m\alpha m\alpha} &\ll & \hbar\omega_{m\alpha},
\nonumber\\
h_{mmmm} + h_{\alpha\alpha\alpha\alpha} - 2h_{mm\alpha\alpha} &\ll & \hbar\omega_{m\alpha}
\label{RPA_validity}
\end{eqnarray}
for all pairs of the orbitals $m,\alpha$. The first of these conditions requires weak exchange interaction
between the environment orbitals, as expected. 
The second condition requires that the exchange interaction between
the rotated orbital wave functions $(\psi_\alpha \pm \psi_m)/\sqrt{2}$ should also be small.
In Appendix \ref{Sec_RPA1} we demonstrate the equivalence between the Hamiltonian (\ref{HRPA}) 
and the usual formulation of RPA approximation adopted in quantum chemistry \cite{Dunning}.

To obtain the correlation energy of the environment orbitals, one needs to find
the frequencies of the normal modes of the Hamiltonian (\ref{HRPA}), $\Omega_{m\alpha}$,  from the equation
\begin{eqnarray}
\det \left[M - \Omega^2\delta_{m\alpha,n\beta}\right] = 0,
\label{Omega}
\end{eqnarray}
where the matrix elements of the matrix $M$ are
\begin{eqnarray}
M_{m\alpha,n\beta} = \omega_{m\alpha}^2 \delta_{m\alpha,n\beta} + \frac{4\sqrt{\omega_{m\alpha}\omega_{n\beta}}}{\hbar} h_{m\alpha n\beta}.
\label{M}
\end{eqnarray}
We use the same indices $m\alpha$ for the frequencies of both uncoupled ($\omega_{m\alpha}$) and coupled ($\Omega_{m\alpha}$)
modes assuming that $\Omega_{m\alpha}\to \omega_{m\alpha}$ in the weak interaction limit $h_{m\alpha n\beta}\to 0$.
Having found $\Omega_{m\alpha}$, one can express the ground state energy of the environment Hamiltonian (\ref{HRPA})  as 
\begin{eqnarray}
E_{\rm env}^{\rm RPA} = E_{\rm env}^{\rm HF} + E_{\rm corr}^{\rm RPA}, 
\end{eqnarray}
where $E_{\rm corr}^{\rm RPA}$ is the correlation energy in the direct RPA approximation \cite{Furche2}
\begin{eqnarray}
E_{\rm corr}^{\rm RPA} = \frac{1}{2}\sum_{m\alpha} \left(\hbar\Omega_{m\alpha} - \hbar\omega_{m\alpha} - 2h_{m\alpha m\alpha} \right).
\label{E_corr}
\end{eqnarray}

Although the Hamiltonian (\ref{H11}) is simpler than the original Hamiltonian (\ref{H}), 
it still cannot be solved exactly.
Therefore, we make yet another approximation and consider the static screening limit of the RPA approximation.
This stronger approximation requires the singlet-triplet gap frequency $\Delta E_{\rm ST}/\hbar$ to be much smaller than
the lowest of the frequencies of the environment oscillators (\ref{omega}).
In Appendix \ref{static} we describe the static limit of RPA in detail
and show that in this limit the system Hamiltonian (\ref{H11}) can be reduced to the form
\begin{eqnarray}
H = E_{\rm env}^{\rm HF} + E_{\rm corr}^{\rm RPA} + \tilde H_{12},
\label{H_static}
\end{eqnarray}
where the modified two-orbital Hamiltonian 
\begin{eqnarray}
\tilde H_{12} = \sum_\sigma\sum_{pq=1}^2 \tilde t_{pq} c^\dagger_{p\sigma} c_{q\sigma}
+  \sum_{\sigma\sigma'} \sum_{pqrs=1}^2 \frac{\tilde h_{psqr}}{2} c^\dagger_{p\sigma} c^\dagger_{q\sigma'} c_{r\sigma'} c_{s\sigma}
\label{H12_mod}
\end{eqnarray}
has the same form as the Hamiltonian (\ref{H012}), but
contains the screened Coulomb integrals $\tilde h_{psqr}$ instead of the non-screened ones $h_{psqr}$. The screened integrals $\tilde h_{psqr}$ read
\begin{eqnarray}
\tilde h_{psqr} = h_{psqr} - 4\sum_{n\beta,k\gamma} h_{ps n\beta} (A+B)^{-1}_{n\beta,k\gamma} h_{k\gamma qr},
\label{h_screened} 
\end{eqnarray}
where the matrices $A$ and $B$ are familiar from the usual formulation of RPA approximation in chemistry \cite{Dunning},
\begin{eqnarray}
A_{m\alpha,n\beta} &=& \hbar\omega_{m\alpha}\delta_{m\alpha,n\beta} + 2 h_{m\alpha n\beta},
\nonumber\\
B_{m\alpha,n\beta} &=& 2 h_{m\alpha n\beta}.
\label{AB}
\end{eqnarray}
We emphasize that  
the exchange corrections do not appear in the matrices $A$ and $B$ because here we consider the direct RPA.

The eigenenergies of the Hamiltonian (\ref{H_static}) are given by the same Eqs. (\ref{Etr},\ref{Esg}), in which 
one should substitute the renormalized parameters defined in the following way:
\begin{eqnarray}
&& U_1 = \frac{\tilde h_{1111}}{2}, \;\; U_2 = \frac{\tilde h_{2222}}{2}, \;\; J_{12} = \tilde h_{1122}, \;\; K_{12} = \tilde h_{1212},
\nonumber\\ &&
t_1 = t'_{12} + \tilde h_{1112},\;\; {\rm and} \;\; t_2 = t'_{12} + \tilde h_{1222}.
\label{UJK_RPA} 
\end{eqnarray}
However, one should keep the original non-screened integrals $h_{psqr}$ (\ref{h_int}) in the hopping matrix elements $t'_{pq}$ (\ref{tpq})
and in the energies of the orbitals $\varepsilon_1 = t'_{11}$, $\varepsilon_2 = t'_{22}$ because they
originate from the Hartree-Fock corrections derived before the RPA approximation has been performed.
The singlet-triplet gap  $\Delta E_{\rm ST}$ is given by Eq. (\ref{dEST}) with the parameters defined by Eqs. (\ref{UJK_RPA}).

 %
 %
\begin{figure}
\begin{tabular}{ccc}
$p-$benzyne & $m-$benzyne & $o-$benzyne \\
\includegraphics[width=2.5cm]{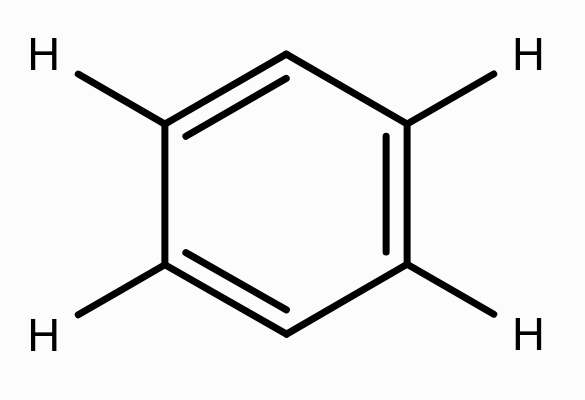} & \includegraphics[width=2.5cm]{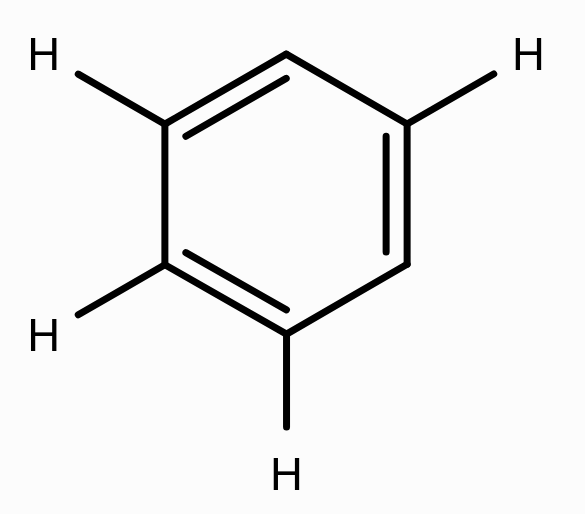} & \includegraphics[width=2.5cm]{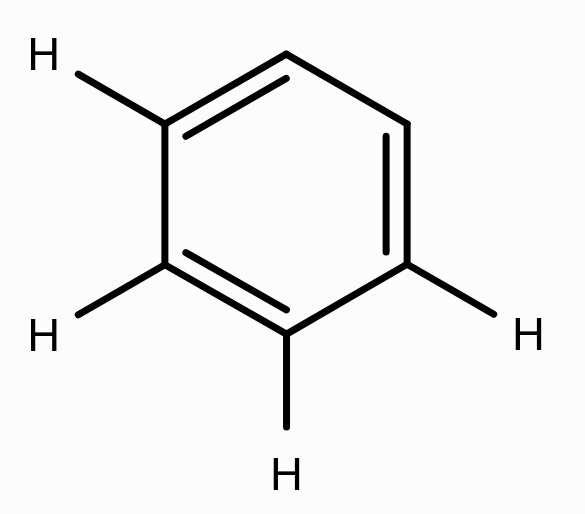} \\
 & & \\
DDP-1 & DDP-2 & TME \\
\includegraphics[width=2.5cm]{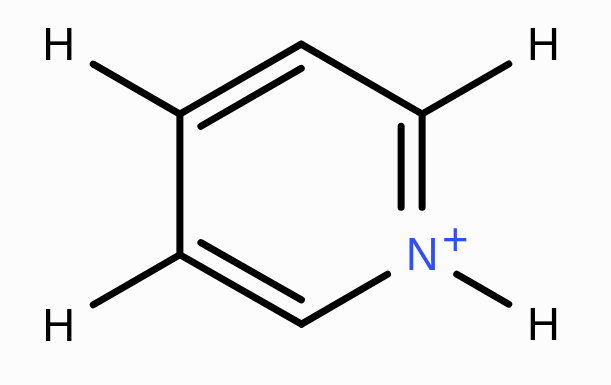} & \includegraphics[width=2.5cm]{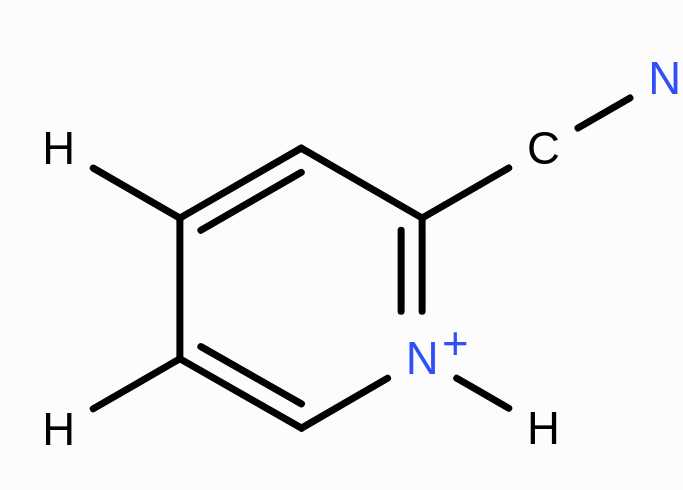} & \includegraphics[width=2.5cm]{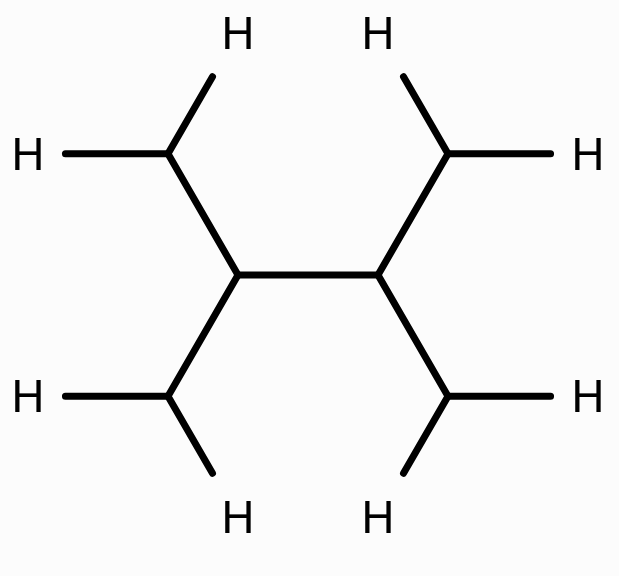} \\
 & & \\
CPC & PN & PC  \\
\includegraphics[width=2.5cm]{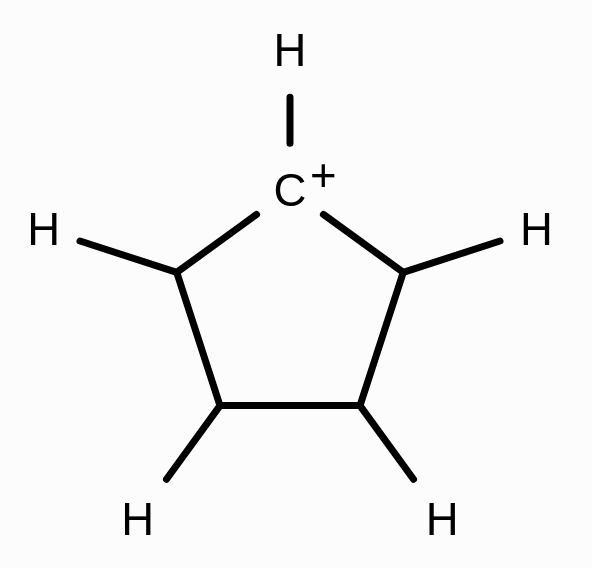} & \includegraphics[width=2.5cm]{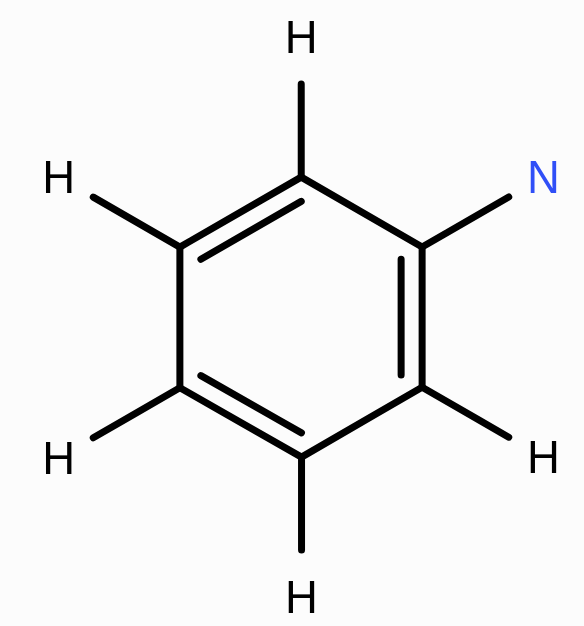} & \includegraphics[width=2.5cm]{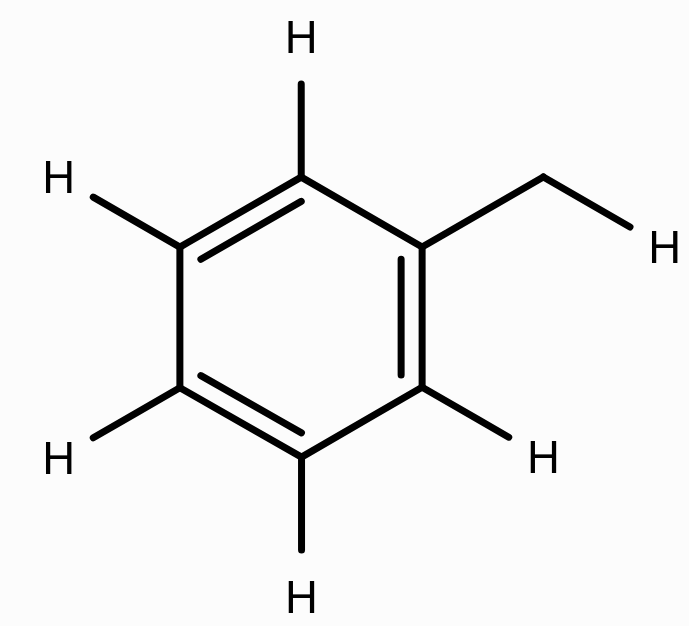} \\
 & TMM &  \\
 & \includegraphics[width=2.5cm]{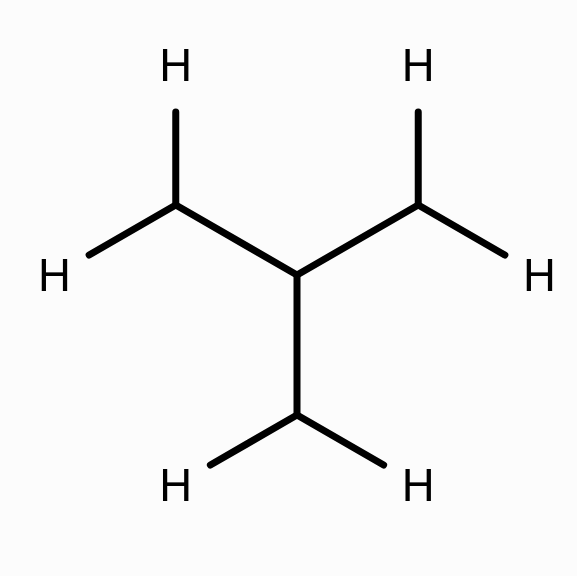} &  \\
\end{tabular}
\caption{Ten molecules considered in our study: (1) $p-$benzyne, (2) $m-$benzyne, (3) $o-$benzyne, (4) 2,5-didehydropyridinium cation (DDP-1) \cite{Sheng}, (5) 6-cyano-2,5-didehydropyridinium cation (DDP2-2) \cite{Sheng},
(6) trimethylethylene (TME), (7) cyclopentane cation  (CPC),
(8) phenylnitrene (PN), (9) phenylcarbene (PC), (10) trimethylenemethane (TMM).}
\label{molecules}
\end{figure}

%
%
\section{Computational details}

\subsection{Molecular test set}

We have tested the static direct RPA approximation for several diradical 
and diradicaloid molecules by comparing its predictions with those of CASSCF and NEVPT2 simulations. 
For this test we have chosen ten molecules shown in Fig. \ref{molecules}.
Six of them have singlet ground state, namely,  
$p$-benzyne, $m$-benzyne, $o$-benzyne, two pyridinium-based benzynes analogues (DDP-1 and DDP-2) \cite{Sheng}, and trimethylethylene (TME).
The remaining four molecules, cyclopentane cation (CPC), phenylnitrene (PN), phenylcarbene (PC) and trimethylenemethane (TMM),
have the triplet ground state.  
The results are presented in Table \ref{results}, where we provide the values of the singlet-triplet splitting $\Delta E_{\rm ST}$
obtained with various methods.

\subsection{Quantum-chemical calculations}
CASSCF \cite{Sun} calculations were carried out with the \textbf{PySCF} package version 2 \cite{Sun1,Sun2,Sun3}. 
All calculations were done using triple-$\zeta$ Def2-TZVP basis set \cite{Schafer} with the default auxillary density-fitting basis.
The active space of the CASSCF calculations was comprised of all $\pi$-orbitals and the non-bonding orbitals 
(shown in the Supporting Information, see p. 12 below). 
The initial orbitals used for CASSCF calculations were UHF natural orbitals \cite{Pulay}.
The active orbitals are shown in the supporting information.
The \textbf{Geometric} software package \cite{Wang2} has been used for geometry optimization. 
All structures were optimized for the ground-state using CASSCF method as opposed to optimizing it for a specified spin state.
Therefore, the singlet-triplet splittings provided in Table \ref{results} are computed at the fixed ground state geometry.
Density fitting was used for optimization calculations \cite{Li}.
After the optimizations, the NEVPT2 calculations have been carried out on top of 
the CASSCF optimized orbitals averaged (SA) over the singlet and the triplet states (SA-CASSCF). 
The one-electron and the two-electron reduced density matrices
have been extracted from the same SA-CASSCF calculations.
To test the predictions of the simple two orbital model, we have also performed the 
CASSCF(2,2) and complete active space configuration interaction (CASCI(2,2)) simulations 
with only two canonical frontier orbitals in the active space.

\section{Results and discussion}
\label{Sec_results}

\begin{table*}[ht]
\begin{tabular}{|c|c|c|c|c|c|c|c|c|c|c|c|}
\hline
                           & $p$-benzyne      & $m$-benzyne      & $o$-benzyne     & DDP-1        & DDP-2    &  TME    & CPC     &  PN      & PC      &  TMM    &  Error \\
                           &                  &                  &                 &              &          &         &         &          &         &         &    \\
\hline
NEVPT2                     & -3.61            & -25.75           & -54.00          & -3.33        & -2.71    & -4.04   &  14.86  &  20.21   & 26.56   & 23.50   &  0  \\
\hline
CASSCF                     & -3.62            & -22.99           & -50.74          & -3.04        & -2.28    & -4.05   &  16.00  &  21.03   & 31.53   & 24.66   &  1.29 \\
\hline
NEVPT2(2,2)                & -2.59            & -20.31           & -44.37          & -1.26        & -0.96    & -2.05   &  28.77  &  23.93   & 19.71   & 72.03   &  9.44 \\
\hline
CASSCF(2,2)                & -1.42            & -19.54           & -43.21          & -1.23        & -0.93    & -2.17   &  19.85  &  30.45   & 30.50   & 11.34   &  6.00  \\
\hline
two orbitals model         & -1.42            & -20.37           & -44.50          & -1.27        & -0.93    & -1.94   &  27.82  &  23.79   & 30.50   & 66.07   &  8.61 \\
Eqs. (\ref{dEST},\ref{UJK}),
triplet                    &                  &                  &                 &              &          &         &         &          &         &         &    \\
\hline
two orbitals model         & -1.42            & -20.38           & -44.58          & -1.28        & -0.94    & -1.88   &  17.53  &  10.73   & 30.50   & 3.76    &  5.88  \\
Eqs. (\ref{dEST},\ref{UJK}),
singlet                    &                  &                  &                 &              &          &         &         &          &         &         &    \\
\hline
RPA, Eqs. (\ref{dEST},
\ref{UJK_RPA}),            & -4.10$^a$        & -29.59           & -57.64          & -2.94        & -2.20    & -2.41   &  20.14  &  23.79   & 22.19$^b$ & 66.07   & 6.63  \\
triplet                    &                  &                  &                 &              &          &         &         &          &         &         &    \\
\hline
RPA, Eqs. (\ref{dEST},
\ref{UJK_RPA}),            & -4.10$^a$        & -29.62           & -57.73          & -2.97        & -2.15    & -2.38   &  13.07  &  10.73   & 22.19$^b$ & 3.76    & 4.6  \\
singlet                    &                  &                  &                 &              &          &         &         &          &         &         &    \\
\hline
$|\Delta E_{\rm ST}|
/\Delta\varepsilon_{\min}$ & 0.0041           &  0.0691          &  0.1667         &  0.0043      &  0.0040  &  0.0057 &  0.0599 &  0.0423  & 0.0516  &  0.1548 &   \\
\hline
\end{tabular}
\caption{Siglet-triplet splitting (kcal/mol) for several molecules. 
The number of orbitals in the active space for CASSCF and NEVPT2 simulations has been chosen as follows:
8 orbitals for $p-$benzyne, $m-$benzyne, $o-$benzyne, DDP-1, PN and PC; 9 orbitals for DDP-2;  6 orbitals for TME; 5 orbitals for CPC; and 4 orbitals for TMM.
The last row shows the ratio of the singlet-triplet gap 
to the minimum enegry splitting between the empty and the doubly occupied orbitals $\Delta\varepsilon_{\min}$.
The last column "Error" shows the absolute value of the error (kcal/mol) relative to NEVPT2 method averaged over all ten molecules. \\
$^a$ For this simulations single and two electron integrals for 104 orbitals obtained from CASSCF(2,2) have been used. \\
$^b$ For this simulations single and two electron integrals for 128 orbitals obtained from CASSCF(2,2) have been used. }
\label{results}
\end{table*}

%
%
\begin{figure*}
	\includegraphics[width=18cm]{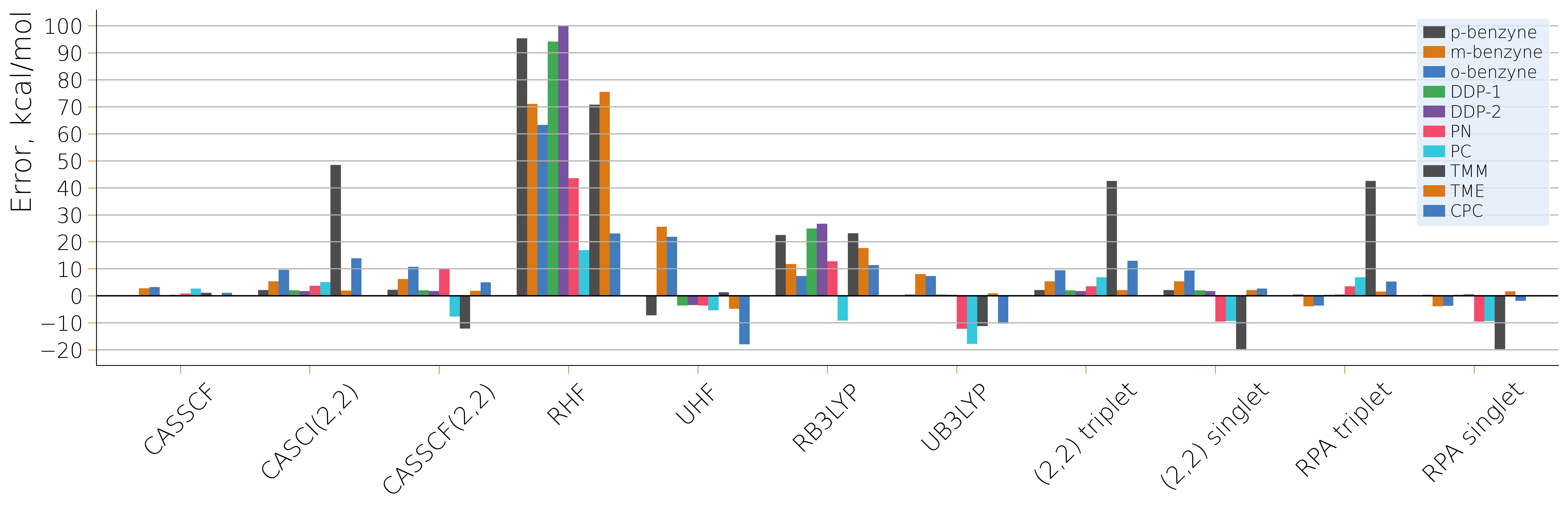}
	\caption{Errors of considered methods in $E_{ST}$ with respect to  the NEVPT2 reference for the molecular test set.}
\label{errors}
\end{figure*}

As an output of the CASSCF simulations we have obtained the 
values of the singlet-triplet splitting $\Delta E_{\rm ST}$, the hopping amplitudes between the orbitals $t_{pq}$ (\ref{t0pq}), 
the Coulomb integrals $h_{psqr}$ (\ref{h_int}) and the single-electron reduced density matrices 
for the singlet, $\rho_{rs}^{\rm sg}$, and for the triplet, $\rho_{rs}^{\rm tr}$, states of all molecules. 
Perfoming CASSCF(2,2) simulations with two orbitals in the active space for $p-$benzyne and for PC molecule
we have generated large sets of the single electron (\ref{t0pq}) and of the two electron (\ref{h_int}) integrals
spanning 104 orbitals for $p-$benzyne and 128 orbitals for PC. All these integrals have been subsequently used 
for the RPA approximation. For the eight remaining molecules we have  stored only one and two-electron intergals corresponding 
to the orbitals of the active space and have tested the RPA model with this limited input.

Next, we have rotated the basis of the active space orbitals in such a way that
the density matrices $\rho_{rs}^{\rm sg}$ and $\rho_{rs}^{\rm tr}$ acquired the diagonal form (\ref{rho}).
The integrals $t_{pq}$ and $h_{psqr}$ have been modified in accordance with this rotation.
The basis rotation has been
performed because the RPA approximation requires the diagonal 
shape of the reduced density matrix for the environment orbitals, while the CASSCF density matrices
with more than two orbitals in the active space are not diagonal.
After these steps, we have used the two-orbital model with the bare parameters (\ref{UJK})
and with the RPA corrected parameters (\ref{UJK_RPA})
to evaluate the singlet-triplet gaps $\Delta E_{\rm ST}$ (\ref{dEST}).
Since the singlet and triplet reduced denstity matrices sometimes require different basis rotations for
their diagonalization, different values of the splitting $\Delta E_{\rm ST}$ are obtained
depending on which density matrix, $\rho_{rs}^{\rm sg}$ or $\rho_{rs}^{\rm tr}$, was chosen.
However, the two resulting values of the gap $\Delta E_{\rm ST}$, given in Table \ref{results}, 
almost coincide for the molecules with the singlet ground state ($\Delta E_{\rm ST}<0$) and for PC both with and without RPA corrections.
This reaffirms the validity of our model for these molecules because the value of $\Delta E_{\rm ST}$ should not depend on the basis
choice.

At the same time, the two values of the gap differ significantly for the three molecules with
the triplet ground state ($\Delta E_{\rm ST}>0$).
We have verified that for these molecules the results based on the
triplet state density matrix are more reliable because both the hopping matrix $t'_{pq}$ 
and  $\rho_{rs}^{\rm tr}$ are close to the diagonal form before and after the
basis rotation. In contrast, 
one cannot bring the singlet density matrix $\rho_{rs}^{\rm sg}$ and the matrix  $t'_{pq}$
to the diagonal form in the same basis.

The obtained values of the gap $\Delta E_{\rm ST}$, provided in Table I, 
demonstrate that the RPA corrections indeed 
improve the agreement between the predictions of the two-orbital model (\ref{dEST}) and the results of CASSCF and NEVPT2 calculations
with large active space.
For the molecules with very small gap, $|\Delta E_{\rm ST}|\lesssim 0.04\Delta\varepsilon_{\min}$, 
the difference between NEVPT2 and the triplet-basis RPA is less than 20\% (see Fig. \ref{errors}) because   
the validity condition of the static RPA approximation (\ref{condition}) is well satisfied.

For the molecules with larger gap,  $|\Delta E_{\rm ST}|\gtrsim 0.05\Delta\varepsilon_{\min}$,
the results are mixed. For $m$-benzyne and for PC the static RPA does not improve the result
of the simple two-orbital model, for $o$-benzyne the improvement is significant,
for TME and CPC molecules the RPA correction somewhat improves the value of $\Delta E_{\rm ST}$, but
the resulting value is still far off from the NEVPT2 reference. 
For PN and TMM the RPA
approximation does not make any difference because the Coulomb integrals between very few active space orbitals used in the simulation are very small,
and one should include more orbitals in the analysis for the RPA corrections to have any effect.

The last column of Table I shows the average absolute value of the error of the method relative to the NEVPT2 reference result,
\begin{eqnarray}
{\rm Error} = \frac{\sum_{j=1}^{10} \left|\Delta E_{\rm ST}^{(j)} - \Delta E_{\rm ST, NEVPT2}^{(j)}\right|}{10},
\end{eqnarray}
where the index $j$ runs over all ten molecules. Although the errors appear rather big, they are dominated
by a single molecule, TMM. 
If one excludes TMM, the errors are significantly reduced, and the RPA based on the
triplet reduced density matrix becomes the second best method, after CASSCF with large active space, 
with the average error 2.64 kcal/mole.

Finally, Fig. \ref{errors} also lists errors (with respect to the NEVPT2 reference) of the single-determinant methods, 
i.e. Hartree-Fock and density functional theory with the standard hybrid functional B3LYP within restricted and unrestricted spin formalisms (RHF/UHF; RB3LYP/UB3LYP). 
As expected, spin-symmetry breaking in the unrestricted formalism dramatically improves the accuracy of $\Delta E_{\rm ST}$, making UB3LYP values compatible in quality with the RPA approximation. 
However, this is achieved by heavy
spin contamination in the singlet state. Furthermore, numerically converging to a spin-symmetry broken SCF solution for a singlet state is not trivial in practical quantum chemical calculations.

\section{Conclusion}
\label{Sec_conlcusion}

We have investigated how accurately the analytically solvable two-orbital model
can describe the properties of diradical molecules. 
We found that one can significantly improve the accuracy
of this model by incorporating the effect of the doubly occupied and empty 
orbitals via the direct RPA approximation in the static limit.
This approximation leads to the renormalization of the parameters
of the two-orbital model without changing its basic structure.
The physical mechanism behind the static RPA approximation is the
screening of the Coulomb interaction between the two orbitals, which host the two
unpaired electrons of the diradical, by the doubly occupied and empty orbitals.
The static RPA is expected to work well
if the singlet-triplet gap of the diradical is much smaller
than the gap between the highest doubly occupied orbital and the lowest
empty orbital. 
Comparing the results of this approximation 
with accurate NEVPT2 simulations for a set of ten
molecules, we have shown that the RPA corrections significantly
improve the accuracy of the two-orbital model for the diradicals with
small singlet-triplet gap, i.e., for \textit{true} diradicals. 
Our approach can be also used to improve the model description of
double quantum dots made of semiconducting materials.

\section{ACKNOWLEDGMENTS}

This work was supported by the German Federal Ministry of Education and Research, through projects Q-Exa (13N16065) and MANIQU (13N15576)
and by the Federal Ministry for Economic Affairs and Climate Action through the project AQUAS (01MQ22003A).

\appendix

\section{RPA Hamiltonian (\ref{HRPA}) and RPA in quantum chemistry}
\label{Sec_RPA1}

In this section we investigate the relation between the Hamiltonian (\ref{HRPA}) and the
standard formulation of the RPA approximation in quantum chemistry \cite{Dunning}.

We begin by expressing the ladder operators of the normal modes $b_{m\alpha},b_{m\alpha}^\dagger$ 
of the environment Hamiltonian (\ref{HRPA}) in terms of the opertors of the uncoupled oscillators $a_{m\alpha},a_{m\alpha}^\dagger$.
For this purpose, 
we introduce the orthogonal matrix $S$, in which every column is an eigenvector ${\bm e}_{m\alpha}$ of the symmetric matrix $M$ (\ref{M}), 
i.e. $M{\bm e}_{m\alpha} = \Omega^2_{m\alpha} {\bm e}_{m\alpha}$, that is, we define the matrix $S$ as
$ S= ({\bm e}_{m_1\alpha_1}, {\bm e}_{m_2\alpha_2},\dots)$, where the dots stand for all remaining eigenvectors of the matrix $M$.
Next, we define the diagonal matrices $\hat\omega$ and $\hat\Omega$ with the matrix elements 
$\hat\omega_{m\alpha, n\beta} = \omega_{m\alpha} \delta_{m\alpha, n\beta}$
and $\hat\Omega_{m\alpha, n\beta} = \Omega_{m\alpha} \delta_{m\alpha, n\beta}$.
We also group the operators $b_{m\alpha},b_{m\alpha}^\dagger$ and $a_{m\alpha},a_{m\alpha}^\dagger$
in the column vectors ${\bm b},{\bm b}^\dagger$ and ${\bm a},{\bm a}^\dagger$, in which every element
correponds to one pair of the orbitals $m\alpha$. Thus, we treat every pair $m\alpha$ as an index
in the Hilbert space with the dimension $N_{\rm occ}\times N_{\rm emp}$, where $N_{\rm occ}$ is the number
of the doubly occupied orbitals and $N_{\rm emp}$ is the number of the empty ones. 
After that, the operators of the normal modes are expressed as
\begin{eqnarray}
{\bm b} &=& 
\frac{\hat\Omega^{\frac{1}{2}} S^T \hat\omega^{-\frac{1}{2}}({\bm a}^\dagger + {\bm a}) 
-  \hat\Omega^{-\frac{1}{2}} S^T \hat\omega^{\frac{1}{2}}({\bm a}^\dagger - {\bm a})}{2},
\nonumber\\
{\bm b}^\dagger &=& 
\frac{\hat\Omega^{\frac{1}{2}} S^T \hat\omega^{-\frac{1}{2}}({\bm a}^\dagger + {\bm a}) 
+ \hat\Omega^{-\frac{1}{2}} S^T \hat\omega^{\frac{1}{2}}({\bm a}^\dagger - {\bm a})}{2}.
\label{normal}
\end{eqnarray}

Next, we consider the equations of motion for the operators $a_{m\alpha}$ and $a^\dagger_{m\alpha}$, 
$i\hbar \dot a_{m\alpha} = [a_{m\alpha},H_{\rm env}]$
and $i\hbar \dot a_{m\alpha}^\dagger = [a_{m\alpha}^\dagger,H_{\rm env}]$.
Evaluating the commutators in the right hand side of these equations with the environment Hamiltonian (\ref{HRPA}), we write them jointly in the matrix form
\begin{eqnarray}
i\hbar\frac{d}{dt}\left(\begin{array}{c} {\bm a} \\ {\bm a}^\dagger \end{array}\right) =
\left( \begin{array}{cc} A & B \\ -B & - A \end{array} \right)
\left(\begin{array}{c} {\bm a} \\ {\bm a}^\dagger \end{array}\right).
\label{RPA}
\end{eqnarray}
The matrices $A$ and $B$ are defined in Eqs. (\ref{AB}).

We look for possible solutions of Eqs. (\ref{RPA}) in the form 
\begin{eqnarray}
\left(\begin{array}{c} {\bm a} \\ {\bm a}^\dagger \end{array}\right) \to
e^{-i\frac{\Delta Et}{\hbar}}\left(\begin{array}{c} {\bm X} \\ {\bm Y} \end{array}\right),
\end{eqnarray}
where $\Delta E = \hbar\Omega_{m\alpha}$ is the difference between the energies of the excited and of the ground state of the molecule,
and ${\bm X}$, ${\bm Y}$  are the vectors of $c-$numbers. 
After that, Eqs. (\ref{RPA}) acquire the form usually used in quantum chemistry \cite{Dunning},
\begin{eqnarray}
\left( \begin{array}{cc} A & B \\ -B & - A \end{array} \right)
\left(\begin{array}{c} {\bm X} \\ {\bm Y} \end{array}\right)
= \Delta E \left(\begin{array}{c} {\bm X} \\ {\bm Y} \end{array}\right).
\label{RPA1}
\end{eqnarray}
The vectors ${\bm X}$ and ${\bm Y}$ are normalized as follows \cite{Dunning}
\begin{eqnarray}
{\bm X}^T{\bm X} - {\bm Y}^T{\bm Y}=1.
\label{norm}
\end{eqnarray}

Next, in quantum chemistry the operator of the elementary excitation with the energy $\Delta E$ above the ground state and with the spin $\sigma$, 
is expressed as \cite{Dunning}
\begin{eqnarray}
s^\dagger_{\Delta E,\sigma} = \sum_{m\alpha} ( X_{m\alpha} c^\dagger_{m\sigma} c_{\alpha\sigma} - Y_{m\alpha} c^\dagger_{\alpha\sigma} c_{m\sigma} ). 
\label{s}
\end{eqnarray}
Let us find the relation between this operator and the operators of bosonic normal mode excitations $b^\dagger_{m\alpha}$.
To do this, from Eqs. (\ref{RPA1}) we derive the following relations 
\begin{eqnarray}
(A-B)(A+B)({\bm X}+{\bm Y})&=&\Delta E^2({\bm X}+{\bm Y}),
\nonumber\\
(A+B)(A-B)({\bm X}-{\bm Y})&=&\Delta E^2({\bm X}-{\bm Y}). 
\label{eqs}
\end{eqnarray}
The matrices $A,B$ are related to the matrix $M$ (\ref{M}) as follows: $(A-B)(A+B) =\hbar^2 \hat\omega^{1/2} M \hat\omega^{-1/2}$ and
$(A+B)(A-B) =\hbar^2 \hat\omega^{-1/2} M \hat\omega^{1/2}$.
Therefore, Eqs. (\ref{eqs}) imply that the vectors $\hat\omega^{-1/2}({\bm X}+{\bm Y})$ 
and $\hat\omega^{1/2}({\bm X}-{\bm Y})$ are proportional to the same eigenvector ${\bm e}_{m\alpha}$ of the matrix $M$,
which corresponds to its' eigenvalue $\Delta E^2/\hbar^2 = \Omega^2_{m\alpha}$.
Thus we can express these vectros as $\hat\omega^{-1/2}({\bm X}+{\bm Y}) = u_1{\bm e}_{m\alpha}$ and $\hat\omega^{1/2}({\bm X}-{\bm Y}) = u_2{\bm e}_{m\alpha}$,
where $u_1$ and $u_2$ are constants. To find them, we first use 
the normalization condition (\ref{norm}), which leads to the relation $u_1u_2=1$. Furthermore, subtracting the equations (\ref{RPA1}) from each other, 
we obtain the new equation
$(A+B)({\bm X} + {\bm Y})=\Delta E ({\bm X} - {\bm Y})$, which in combination with the matrix relation $A+B=\hbar \hat\omega^{-1/2} M \hat\omega^{-1/2}$
leads to $\Omega_{m\alpha} u_1 = u_2$. From these two conditions we find $u_1 = 1/\sqrt{\Omega_{m\alpha}}$ and $u_2 = \sqrt{\Omega_{m\alpha}}$.
This, in turn, leads to the following relations between the components $n\beta$ of the vectors ${\bm X},{\bm Y}$ and of the eigenvector of the matrix $M$, ${\bm e}_{m\alpha}$ :
\begin{eqnarray}
X_{n\beta} + Y_{n\beta} &=& \sqrt{\frac{\omega_{n\beta}}{\Omega_{m\alpha}}}\, e_{m\alpha}^{n\beta},
\nonumber\\
X_{n\beta} - Y_{n\beta} &=& \sqrt{\frac{\Omega_{m\alpha}}{\omega_{n\beta}}}\, e_{m\alpha}^{n\beta}.
\label{relations}
\end{eqnarray}
Next, we perform the summation over the spin index in Eq. (\ref{s})
and invoke the relation (\ref{rep}) between the electronic and bosonic operators.  
Thus we obtain
\begin{eqnarray}
\frac{s^\dagger_{\hbar\Omega_{m\alpha},\uparrow} + s^\dagger_{\hbar\Omega_{m\alpha},\downarrow}}{\sqrt{2}}
\leftrightarrow \sum_{m\alpha} ( X_{m\alpha}a^\dagger_{m\alpha}  - Y_{m\alpha}a_{m\alpha}  ).
\end{eqnarray}
Rearranging the terms as
\begin{eqnarray}
&& \frac{s^\dagger_{\hbar\Omega_{m\alpha},\uparrow} + s^\dagger_{\hbar\Omega_{m\alpha},\downarrow}}{\sqrt{2}}
\leftrightarrow \frac{1}{2}\sum_{m\alpha} \big[(X_{m\alpha} - Y_{m\alpha})(a^\dagger_{m\alpha} + a_{m\alpha})
\nonumber\\ &&
+\, (X_{m\alpha} + Y_{m\alpha})(a^\dagger_{m\alpha} - a_{m\alpha})\big],
\end{eqnarray}
using the relations (\ref{relations}) and comparing the result with the $m\alpha$ component of Eq. (\ref{normal}), we arrive at
\begin{eqnarray}
\frac{s^\dagger_{\hbar\Omega_{m\alpha},\uparrow} + s^\dagger_{\hbar\Omega_{m\alpha},\downarrow}}{\sqrt{2}} \leftrightarrow b^\dagger_{m\alpha}.
\label{rel}
\end{eqnarray}
This relation points to the equivalence of the RPA Hamiltonian (\ref{HRPA}) and the quantum chemistry
formulation of the RPA approximation based on Eqs. (\ref{RPA1}).
We emphasize that the direct RPA approximation, which we use in this paper, deals only with the charge density operators. Therefore,
in Eqs. (\ref{rep},\ref{rel}) summation over the spin indexes is performed. 
Finally, in Eq. (\ref{rel}) we use the arrow $\leftrightarrow$ to emphasize that the operators
on both sides of this relation act in different Hilbert spaces and,
therefore, cannot be equal to each other.

\section{Static limit of direct RPA}
\label{static}

In this section we consider the static screening limit of the direct RPA approximation. 
The idea behind this can be easily understood if one invokes the formal analogy between the
Hamiltonian (\ref{H11}) and the quantum electrodynamics (QED) Hamiltonian, which describes the interaction between electrons and
electromagnetic field. It is well known that by using path integral approach for the QED problem one can integrate out
the electromangentic fields, leaving only the fermionic fields describing the electrons in the action. 
If one then disregards the relativistic retardation effects caused by the finite value of the speed of light, and
considers the static approximation replacing the exact photon Green's function by its zero frequency limit, 
then the Coulomb interaction term with the normal ordering of the fermionic operators is obtained. 
This term is the last term of the Hamiltonian (\ref{H}).
In this section we carry out the same set of approximations
for the Hamiltonian (\ref{H11}). 
However, instead of using rigorous, but lengthy, path integral approach,
we only briefly sketch the derivation by using semi-quantitative method focusing on the physical meaning of 
the used approximations.     

We begin by introducing the effective coordinates and momenta of the environment oscillators as
\begin{eqnarray}
q_{m\alpha}=\sqrt{\frac{\hbar}{2}}(a^\dagger_{m\alpha} + a_{m\alpha}), \;\;\; p_{m\alpha}=i\sqrt{\frac{\hbar}{2}}(a^\dagger_{m\alpha} - a_{m\alpha}).
\end{eqnarray}
They obey the usual commutation rule, $[p_{m\alpha},q_{m\alpha}]=-i\hbar$.
Omitting for the moment the constant energy shifts, we write the Hamiltonian (\ref{H11}) in the form
\begin{eqnarray}
H &=& H_{12} + \sum_{m\alpha} \frac{\omega_{m\alpha} (p^2_{m\alpha} + q^2_{m\alpha})}{2} + \frac{2}{\hbar}\sum_{m\alpha,n\beta} h_{m\alpha n\beta} q_{m\alpha} q_{n\beta}
\nonumber\\ &&
+\, \frac{2}{\sqrt{\hbar}}\sum_{m\alpha} \sum_\sigma \sum_{p,q=1}^2 h_{pqm\alpha} c^\dagger_{p\sigma} c_{q\sigma} q_{m\alpha}.
\label{H111}
\end{eqnarray}
The equations of motion for the operators $q_{m\alpha}$ and $p_{m\alpha}$ are
\begin{eqnarray}
\dot q_{m\alpha} &=& \frac{i}{\hbar}[H,q_{m\alpha}] = \omega_{m\alpha} p_{m\alpha},
\nonumber\\
\dot p_{m\alpha} &=& \frac{i}{\hbar}[H,p_{m\alpha}] = -\omega_{m\alpha} q_{m\alpha} - \frac{4}{\hbar}\sum_{n\beta} h_{m\alpha n\beta} q_{n\beta}
\nonumber\\ &&
-\, \frac{2}{\sqrt{\hbar}} \sum_\sigma \sum_{p,q=1}^2 h_{pqm\alpha} c^\dagger_{p\sigma} c_{q\sigma}.
\end{eqnarray}
Excluding $p_{m\alpha}$ form these equations, we obtain the set of equations for a number of coupled forced harmonic oscillators in the operator form,
\begin{eqnarray}
\ddot q_{m\alpha} + \omega_{m\alpha}^2 q_{m\alpha} + \frac{4\omega_{m\alpha}}{\hbar}\sum_{n\beta} h_{m\alpha n\beta} q_{n\beta}
\nonumber\\
= - \frac{2\omega_{m\alpha}}{\sqrt{\hbar}} \sum_\sigma\sum_{p,q=1}^2 h_{pq m\alpha} c_{p\sigma}^\dagger c_{q\sigma}.
\label{osc}
\end{eqnarray}
In the static screening approximation one can split the coordinates $q_{m\alpha}$ into the static part $q_{m\alpha}^{(0)}$ and the fluctuating part
$\delta q_{m\alpha}$, which approximately commute with each other,
\begin{eqnarray}
q_{m\alpha} = q_{m\alpha}^{(0)} + \delta q_{m\alpha}, \;\;\; [q_{m\alpha}^{(0)},\delta q_{n\beta}] \approx 0.
\label{split}
\end{eqnarray} 
The fluctuating part $\delta q_{m\alpha}$ satisfies the free equation,
\begin{eqnarray}
\delta\ddot q_{m\alpha} + \omega_{m\alpha}^2 \delta q_{m\alpha} + \frac{4\omega_{m\alpha}}{\hbar}\sum_{n\beta} h_{m\alpha n\beta} \delta q_{n\beta} = 0,
\end{eqnarray}
while the static part satisfies Eq. (\ref{osc}) with the second time derivative omitted, 
\begin{eqnarray}
&& \omega_{m\alpha}^2 q_{m\alpha}^{(0)} + \frac{4\omega_{m\alpha}}{\hbar}\sum_{n\beta} h_{m\alpha n\beta} q_{n\beta}^{(0)}
\nonumber\\ &&
= - \frac{2\omega_{m\alpha}}{\sqrt{\hbar}} \sum_\sigma\sum_{p,q=1}^2 h_{pq m\alpha} c_{p\sigma}^\dagger c_{q\sigma}.
\end{eqnarray}
This equation can be re-written in terms of the matrices $A$ and $B$  (\ref{AB}),
\begin{eqnarray}
(A+B){\bm q}^{(0)} = -\hbar {\bm F},
\label{q0}
\end{eqnarray} 
where the vector ${\bm F}$ is defined as
\begin{eqnarray}
{\bm F} = \frac{2}{\sqrt{\hbar}} \sum_\sigma\sum_{p,q=1}^2 h_{pq m\alpha} c_{p\sigma}^\dagger c_{q\sigma}. 
\label{F}
\end{eqnarray}

The static screening approximation outlined above is justified if the oscillator coordinates $q_{m\alpha}$ are fast, while the 
operators characterizing the diradical orbitals 1 and 2, namely, the operators $c_{p\sigma}^\dagger c_{q\sigma}$, are slow. Formally, this condition
can be formulated in the form
\begin{eqnarray}
|\Delta E_{\rm ST}| \ll \Delta\varepsilon_{\min} = \min|\varepsilon_m - \varepsilon_\alpha|,
\label{condition}
\end{eqnarray}
i.e. the energy separation between the diradical orbitals should be much smaller than the 
energy gap between the empty and the doubly occupied orbitals of the environment. The condition (\ref{condition}) is satisfied in good diradicals.

Solving Eq. (\ref{q0}), we obtain the shifts of the oscillator coordinates induced by presence of the diradical orbitals, 
${\bm q}^{(0)} = -\hbar (A+B)^{-1}{\bm F}$. Adding the fluctuating part of the coordinates we obtain
${\bm q} = -\hbar (A+B)^{-1}{\bm F} + \delta {\bm q}$.  Next, we substitute this expression back in the Hamiltonian (\ref{H111}). 
To facilitate this procedure, we first re-write the latter in the more compact form   
\begin{eqnarray}
H = H_{12} + \sum_{m\alpha} \frac{\omega_{m\alpha}p_{m\alpha}^2}{2} + \frac{1}{2\hbar} {\bm q}^T (A+B) {\bm q} + {\bm F}^T{\bm q}.
\end{eqnarray}
After the substitution and the cancelation of certain terms, we split the resulting Hamiltonian into two parts,
\begin{eqnarray}
H = \tilde H_{12} + H_{\rm env}.
\label{Hstatic}
\end{eqnarray} 
The first part in this Hamiltonian can be interpreted as the modified two-orbital Hamiltonian of the orbitals 1 and 2,
\begin{eqnarray}
\tilde H_{12} = H_{12} - \frac{\hbar}{2}  {\bm F}^T (A+B)^{-1} {\bm F},
\label{H1122}
\end{eqnarray}
and the second part --- as the new Hamiltonian of the environment 
\begin{eqnarray}
H_{\rm env} = \sum_{m\alpha} \frac{\omega_{m\alpha}p_{m\alpha}^2}{2} + \frac{1}{2\hbar}  \delta{\bm q}^T (A+B)  \delta{\bm q}.
\label{Henv1}
\end{eqnarray}
Since the shift of coordinate does not change the frequency of an oscillator,
$H_{\rm env}$ (\ref{Henv1}) retains the same form (\ref{HRPA}) if expressed in terms of the new shifted ladder operators
$a_{m\alpha} = \sqrt{2/\hbar}(\delta q_{m\alpha} + i p_{m\alpha})$, $a_{m\alpha}^\dagger = \sqrt{2/\hbar}(\delta q_{m\alpha} - i p_{m\alpha})$. 
We also note that the fermionic operators, which are contained in the vectors ${\bm F}^T$ and ${\bm F}$ of the last
term of the Hamiltonian (\ref{H1122}), are not normally ordered and appear there as the product 
$c_{p\sigma}^\dagger c_{s\sigma} c_{q\sigma'}^\dagger c_{r\sigma'}$. It is an artefact of our simplified treatment of the problem.
As we mentioned above, the rigorous procedure analogous to that in the QED theory, results in the normally ordered
product of the operators. Hence, in the last term of Eq. (\ref{H1122}) we should replace
$c_{p\sigma}^\dagger c_{s\sigma} c_{q\sigma'}^\dagger c_{r\sigma'} \to c_{p\sigma}^\dagger  c_{q\sigma'}^\dagger c_{r\sigma'} c_{s\sigma}$. 
Next, substituting the vector ${\bm F}$ (\ref{F}) in the Hamiltonian (\ref{H1122}), we bring it to the form
(\ref{H12_mod}) given in Sec. \ref{Sec_RPA}.
Finally, since in the Hamiltonian (\ref{Hstatic}) of the environment is no longer interacting with 
the diradical orbitals, we can replace $H_{\rm env}$ by its average value in the ground state
$\langle H_{\rm env}\rangle = E_{\rm corr}^{\rm RPA}$. Restoring the omitted constant energy shift,
we arrive at the final expression for the Hamiltonian (\ref{H_static}).

Physically, the modified Coulomb integrals $\tilde h_{psqr}$ (\ref{h_screened}) represent the matrix elements of the Coulomb potential
screened by presence of the doubly occupied orbitals in the molecule. 
To clarify the physical meaning of Eq. (\ref{h_screened}) even further, we notice that in homogeneous electron gas,
where the basis of plane waves can be used, the integrals (\ref{h_screened}) are proportional to the Fourier component of the
screened Coulomb potential $V_k = -4\pi e^2/k^2\epsilon(0,k)$, where $\epsilon(\omega,k)$ is
the electric susceptibility of the free electron gas in Lindhard approximation \cite{Lindhard}. The appearance
of $\epsilon(0,k)$ instead of $\epsilon(\omega,k)$ in $V_k$ implies the static screening limit $\omega=0$, as we discussed
above. It is known that at sufficiently long distance the screened Coulomb potential of the free electron gas
behaves as $V_{\rm TF}(r) = \int e^{i{\bm k}{\bm r}} (d^3{\bm k}/(2\pi)^3)\;V_k \approx (e^2/r)\, e^{-r/\lambda_{\rm TF}}$. 
Thus, for two orbitals immersed in the homogeneous electron
gas the modified Coulomb integrals (\ref{h_screened}) can be written in the form (\ref{h_int}) with the Coulomb potential $e^2/r$
replaced by $V_{\rm TF}(r)$.



\clearpage


\begin{widetext}

\begin{center}
{\large\bf Supporting Information}
\vspace{0.25cm}

{\bf Active orbitals for the molecular test set}
\end{center}

\begin{figure*}[h]
	\begin{tabular}{cccc}
		\includegraphics[width=3.7cm]{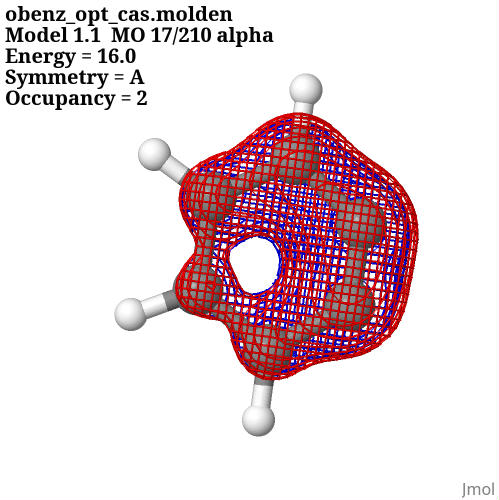} & \includegraphics[width=3.7cm]{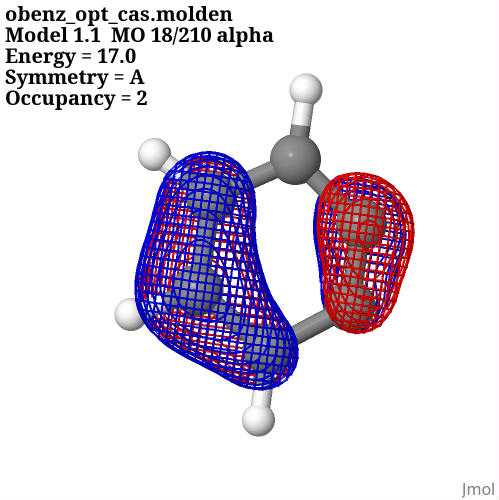} &
		\includegraphics[width=3.7cm]{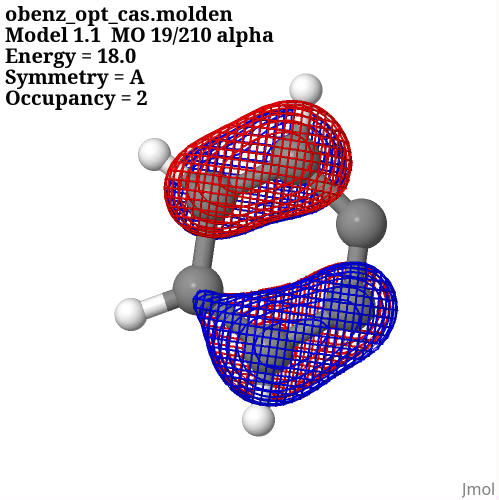} & \includegraphics[width=3.7cm]{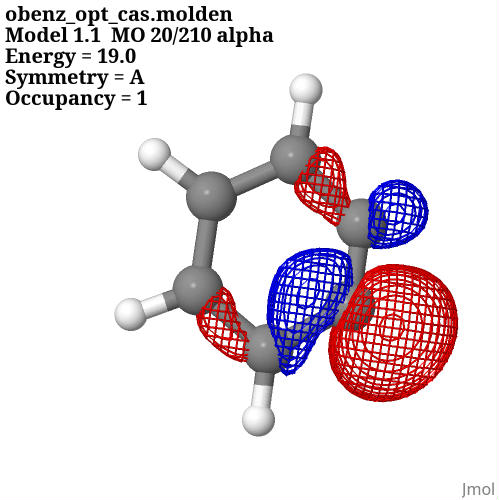} \\
		& & \\
		\includegraphics[width=3.7cm]{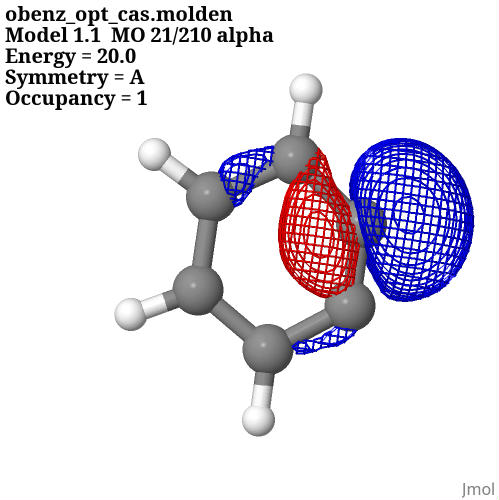} & \includegraphics[width=3.7cm]{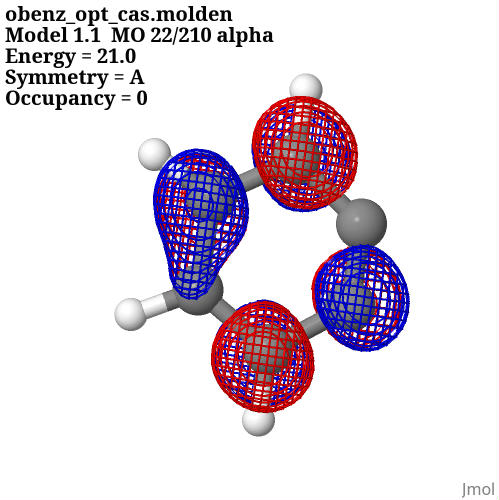} &
		\includegraphics[width=3.7cm]{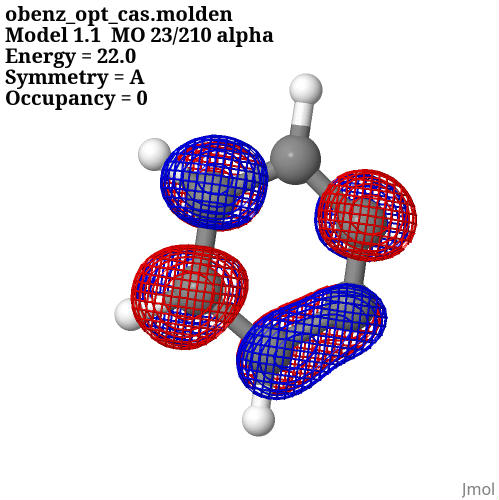} & \includegraphics[width=3.7cm]{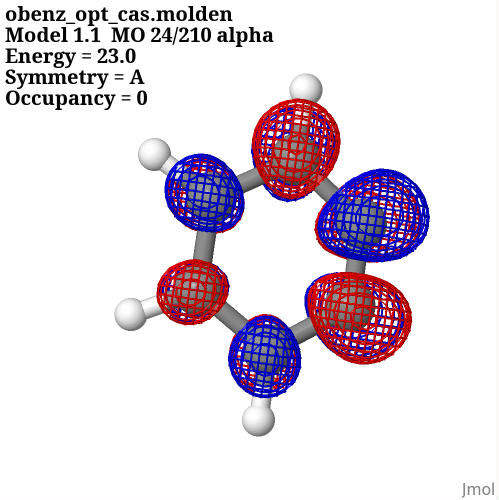} \\
	\end{tabular}
	\caption{$o$-benzyne.}
	\label{molecules}
\end{figure*}

\begin{figure*}[h]
	\begin{tabular}{cccc}
		\includegraphics[width=4cm]{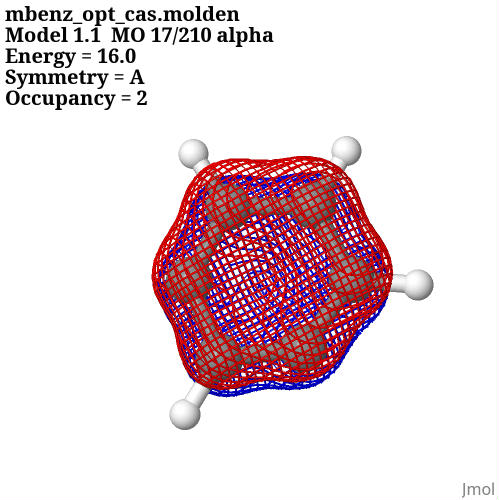} & \includegraphics[width=4cm]{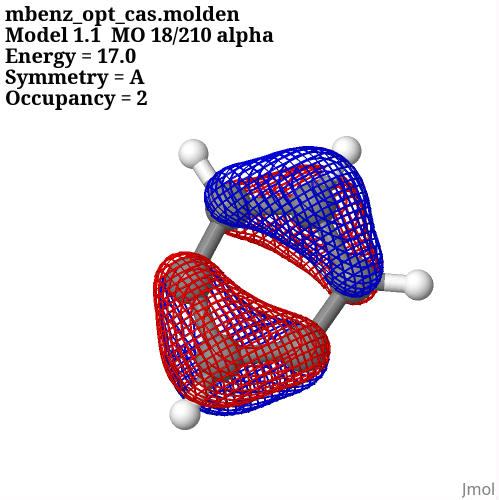} &
		\includegraphics[width=4cm]{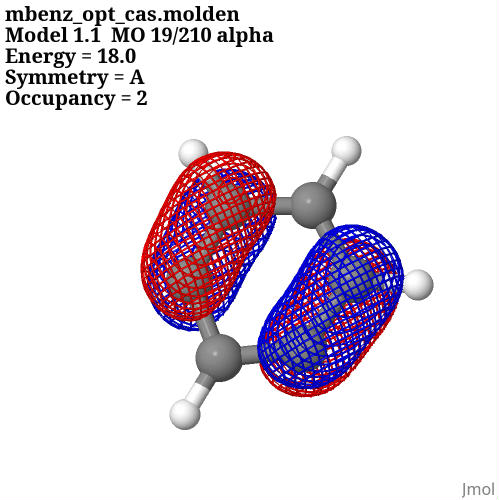} & \includegraphics[width=4cm]{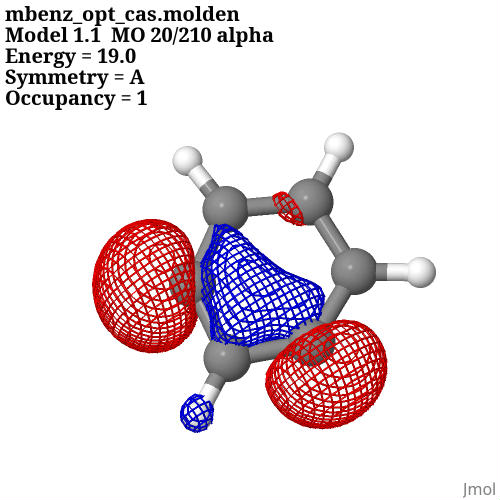} \\
		& & \\
		\includegraphics[width=4cm]{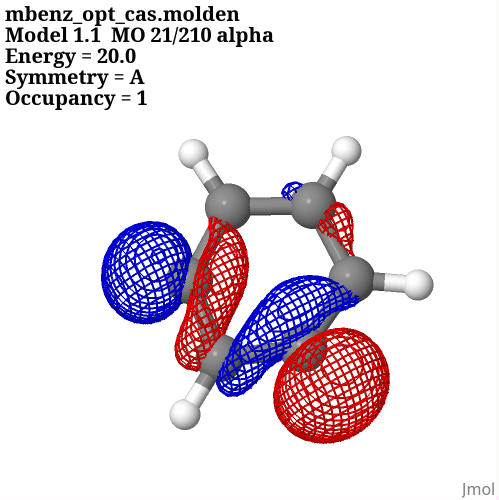} & \includegraphics[width=4cm]{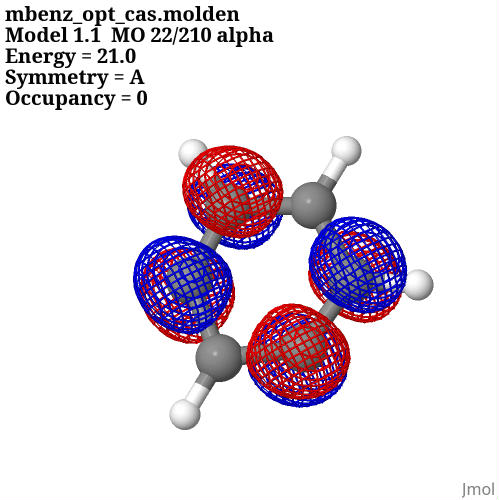} &
		\includegraphics[width=4cm]{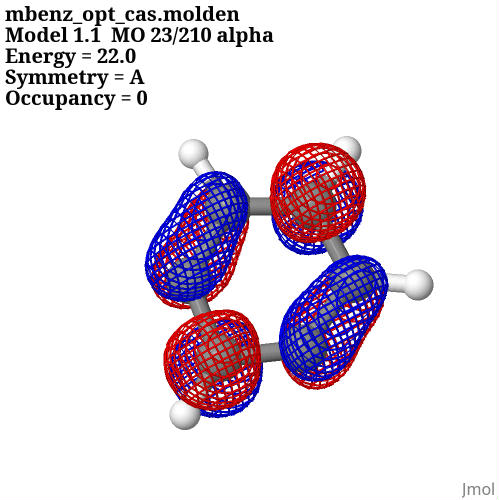} & \includegraphics[width=4cm]{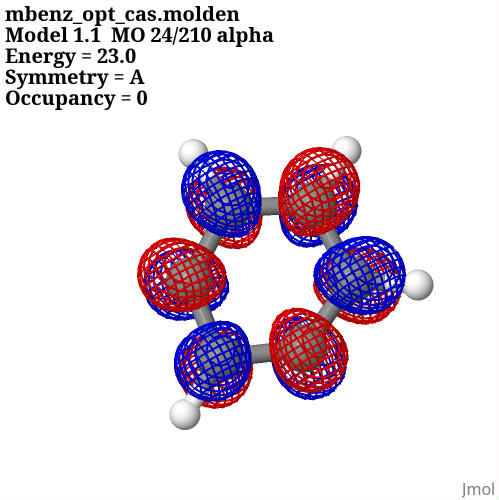} \\
	\end{tabular}
	\caption{$m$-benzyne.}
	\label{molecules}
\end{figure*}

\begin{figure*}[h]
	\begin{tabular}{cccc}
		\includegraphics[width=4cm]{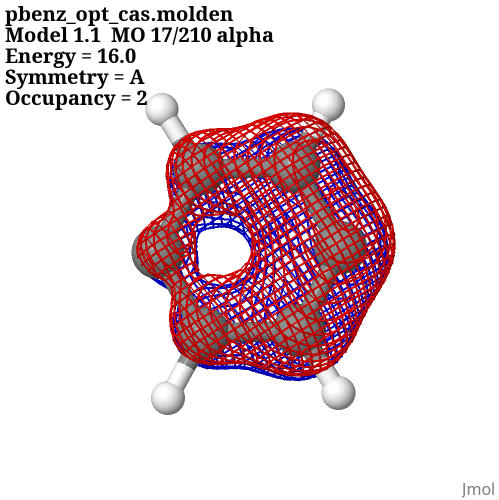} & \includegraphics[width=4cm]{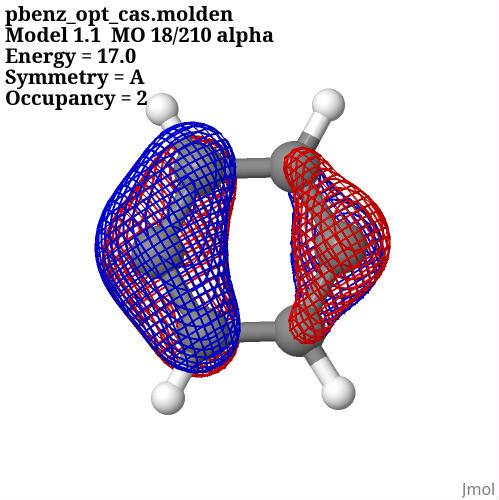} &
		\includegraphics[width=4cm]{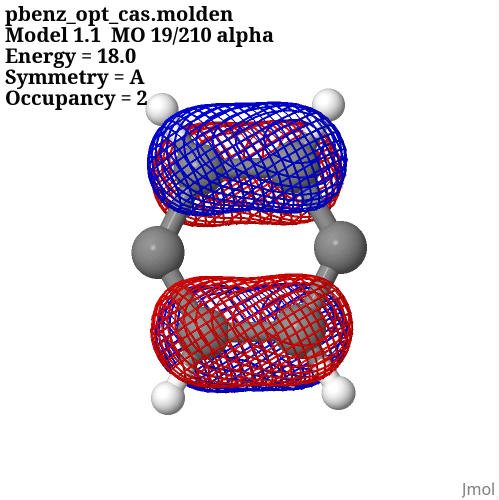} & \includegraphics[width=4cm]{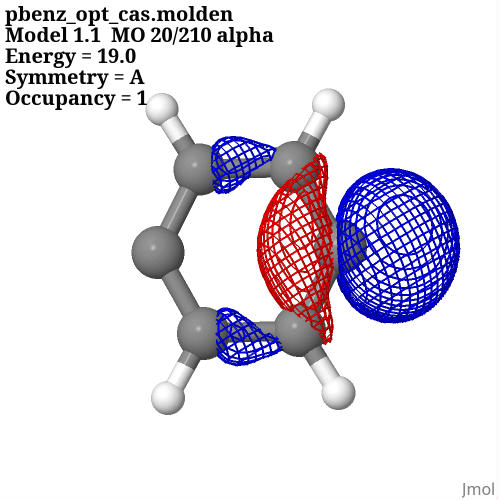} \\
		& & \\
		\includegraphics[width=4cm]{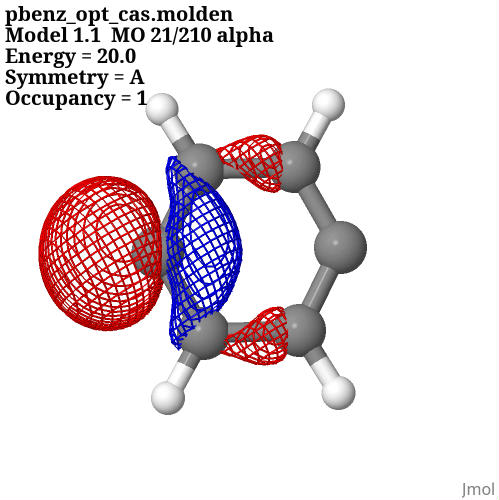} & \includegraphics[width=4cm]{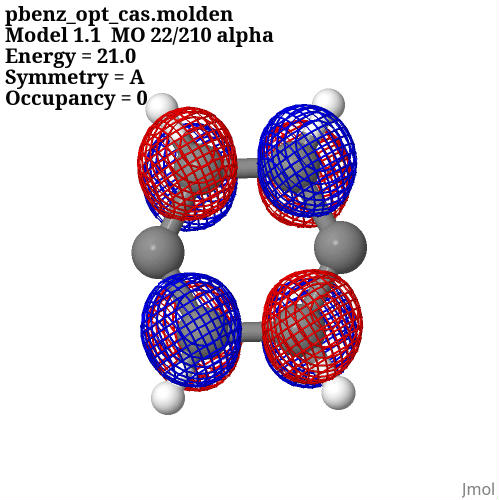} &
		\includegraphics[width=4cm]{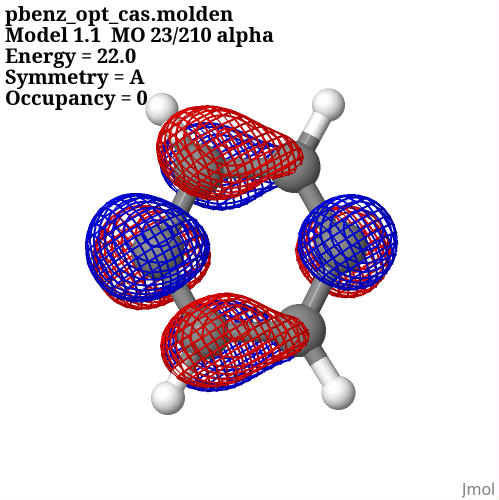} & \includegraphics[width=4cm]{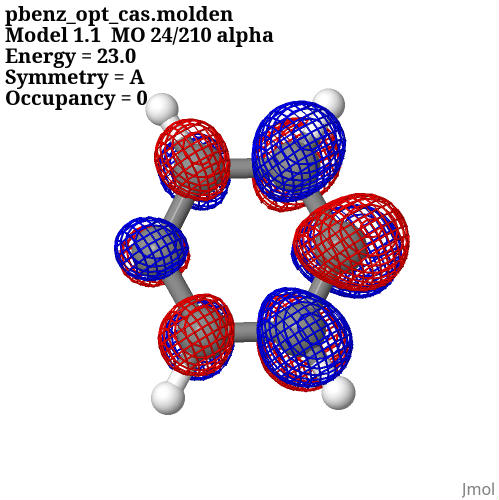} \\
	\end{tabular}
	\caption{$p$-benzyne.}
	\label{molecules}
\end{figure*}

\begin{figure*}[h]
	\begin{tabular}{cccc}
		\includegraphics[width=4cm]{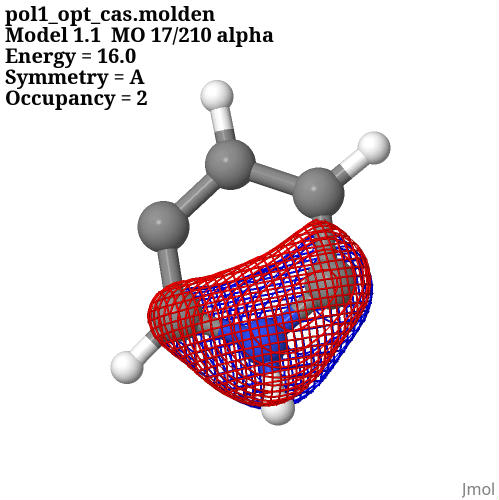} & \includegraphics[width=4cm]{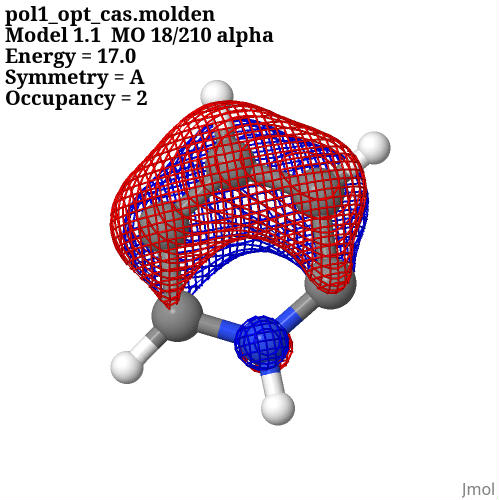} &
		\includegraphics[width=4cm]{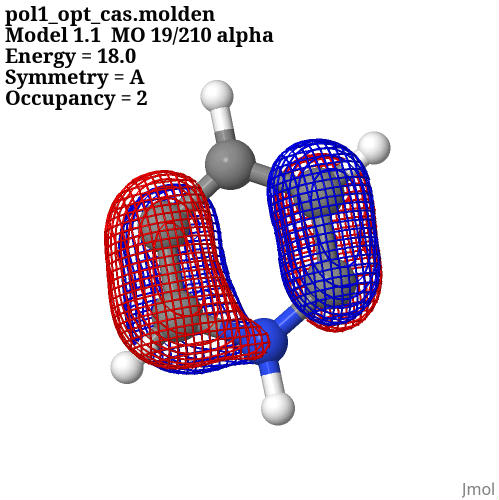} & \includegraphics[width=4cm]{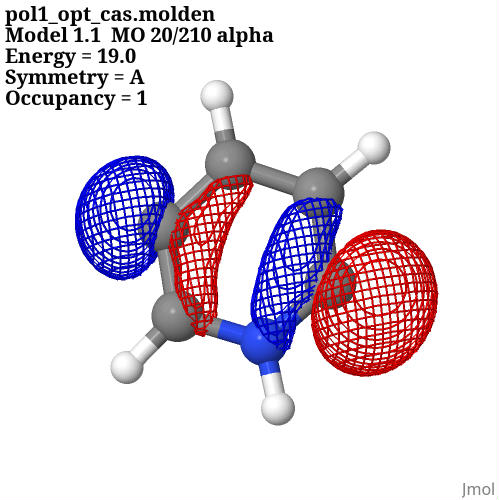} \\
		& & \\
		\includegraphics[width=4cm]{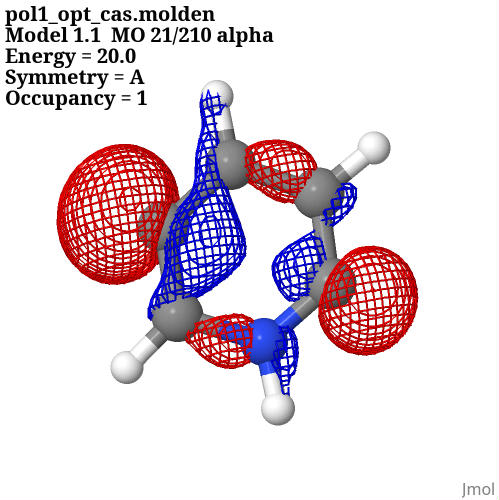} & \includegraphics[width=4cm]{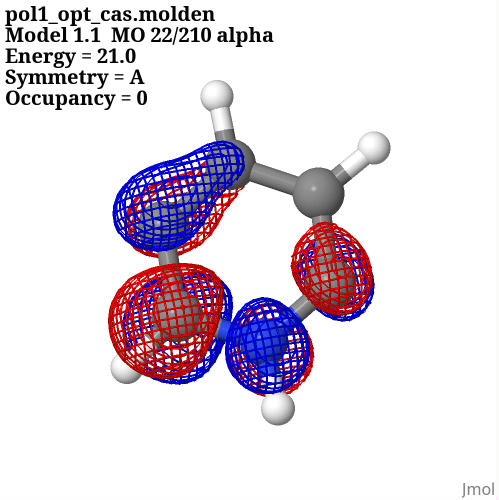} &
		\includegraphics[width=4cm]{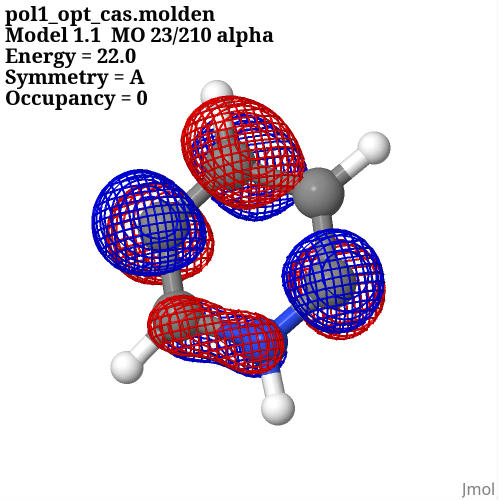} & \includegraphics[width=4cm]{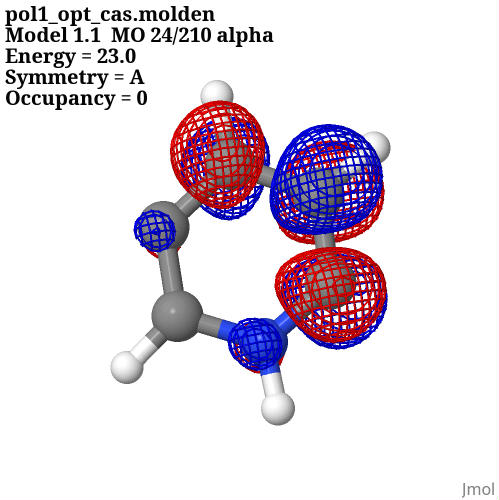} \\
	\end{tabular}
	\caption{2,5-didehydropyridinium cation (DDP-1).}
	\label{molecules}
\end{figure*}

\begin{figure*}[h]
	\begin{tabular}{cccc}
		\includegraphics[width=4cm]{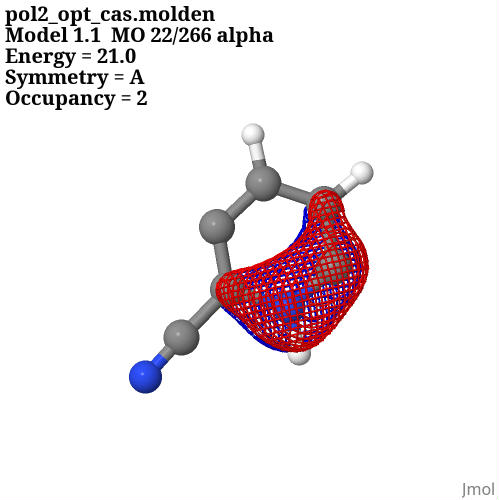} & \includegraphics[width=4cm]{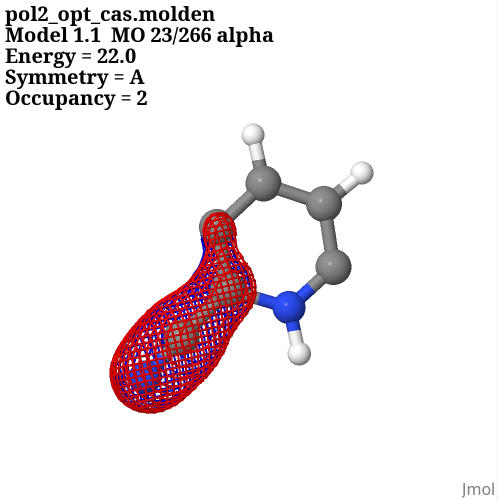} &
		\includegraphics[width=4cm]{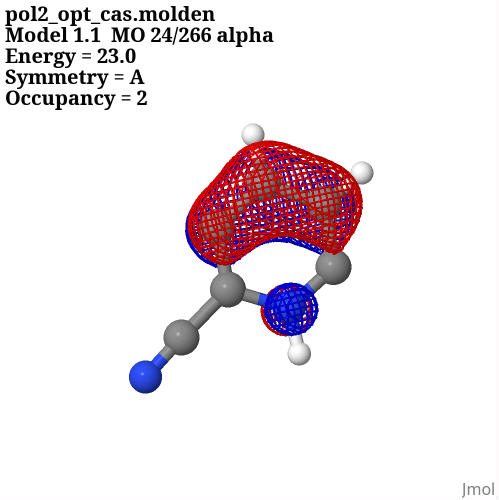} & \includegraphics[width=4cm]{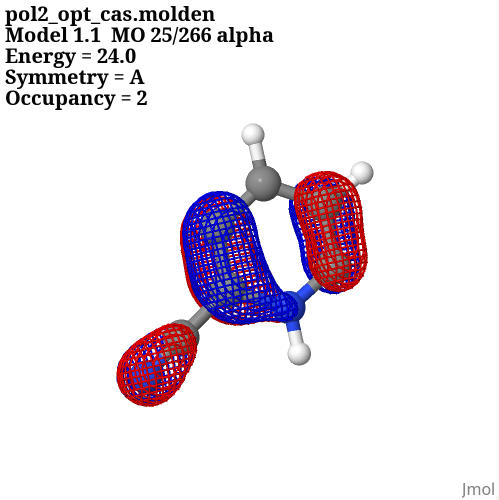} \\
		& & \\
		\includegraphics[width=4cm]{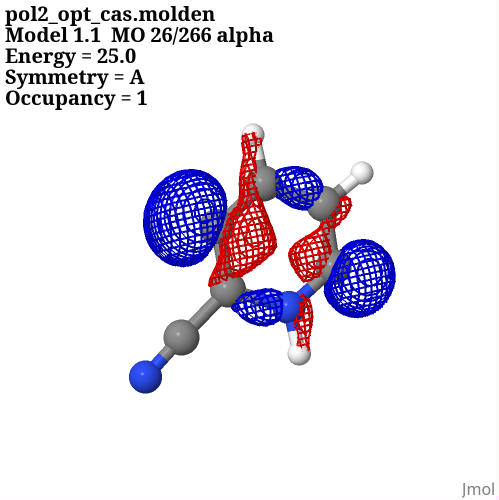} & \includegraphics[width=4cm]{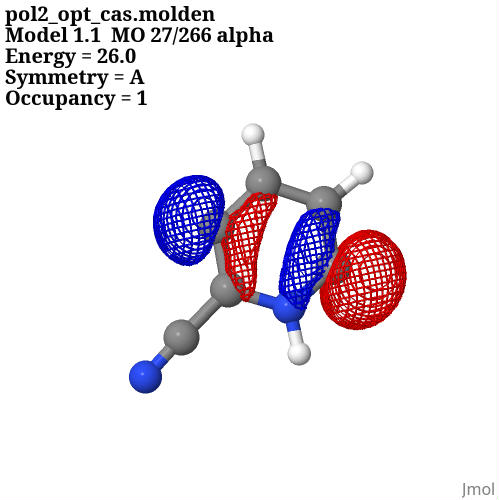} &
		\includegraphics[width=4cm]{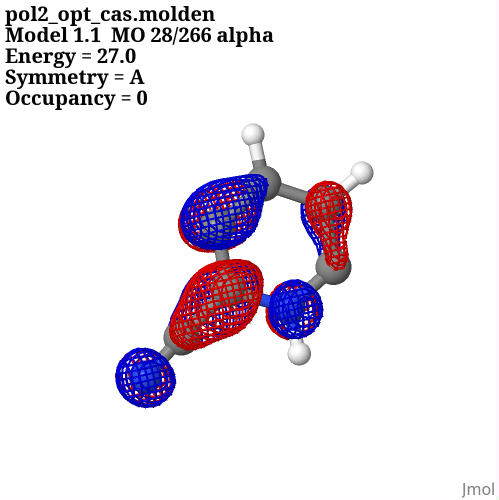} & \includegraphics[width=4cm]{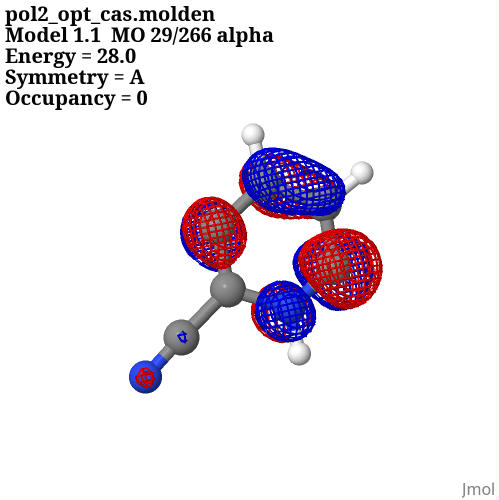} \\
		& & \\
		\includegraphics[width=4cm]{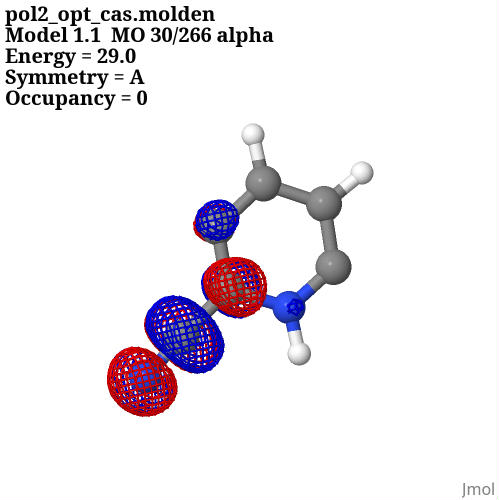} \\
	\end{tabular}
	\caption{6-cyano-2,5-didehydropyridinium cation (DDP2-2)}
	\label{molecules}
\end{figure*}

\begin{figure*}[h]
	\begin{tabular}{ccc}
		\includegraphics[width=4cm]{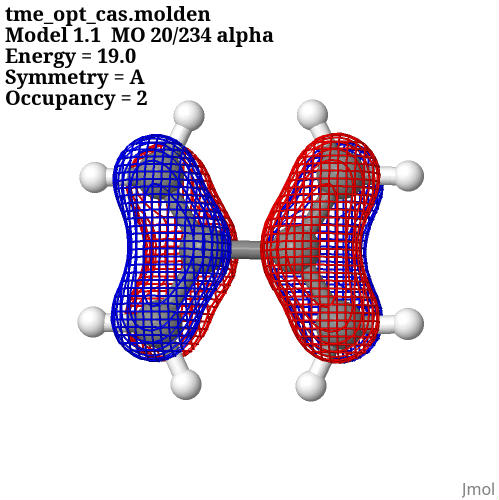} & \includegraphics[width=4cm]{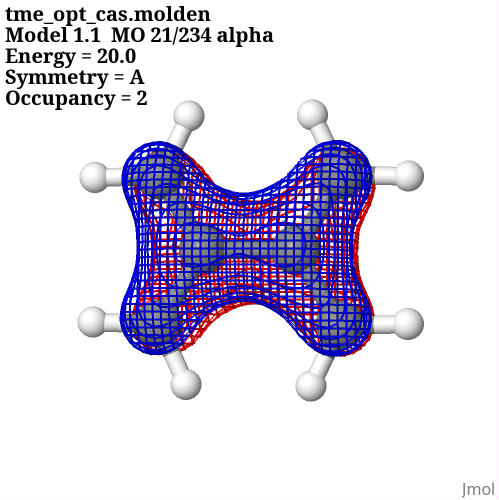} &
		\includegraphics[width=4cm]{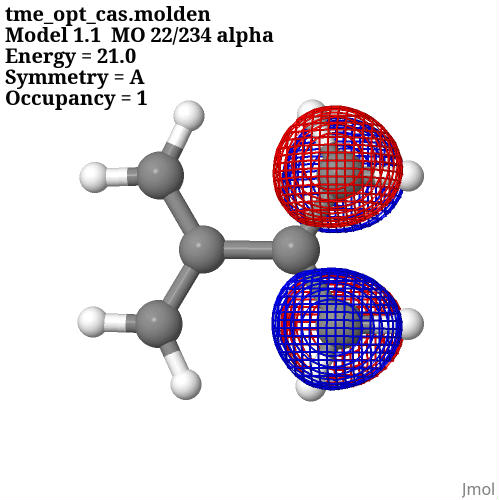} \\
		& & \\
		\includegraphics[width=4cm]{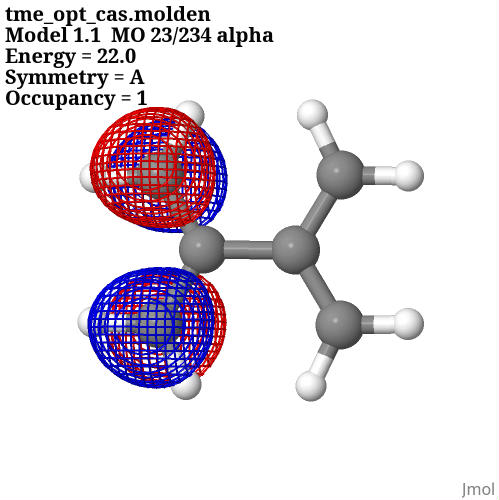} & \includegraphics[width=4cm]{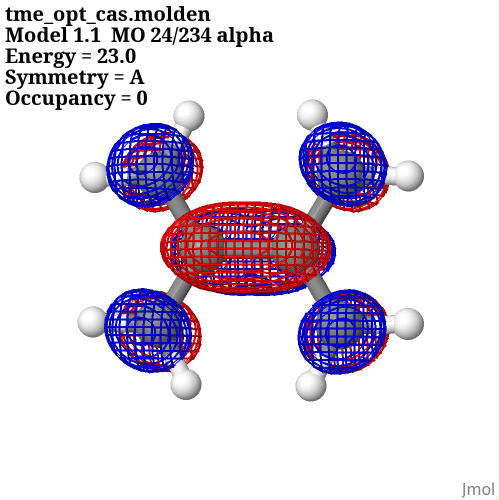} &
		\includegraphics[width=4cm]{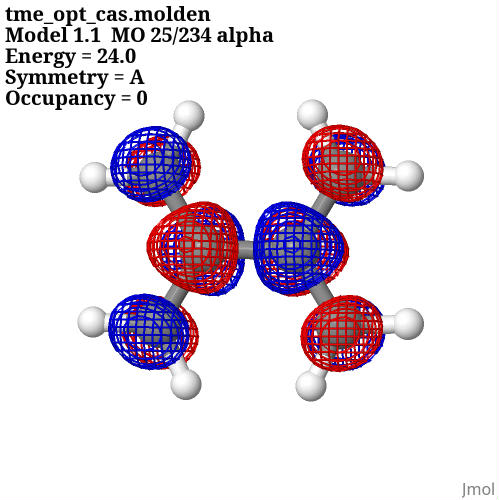} \\
	\end{tabular}
	\caption{trimethylethylene (TME).}
	\label{molecules}
\end{figure*}

\begin{figure*}[h]
	\begin{tabular}{cccc}
		\includegraphics[width=4cm]{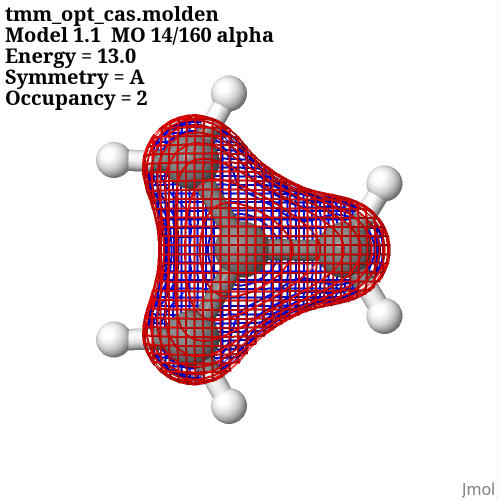} & \includegraphics[width=4cm]{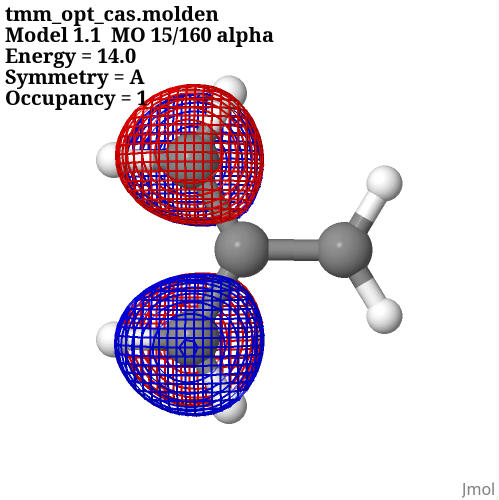} &
		\includegraphics[width=4cm]{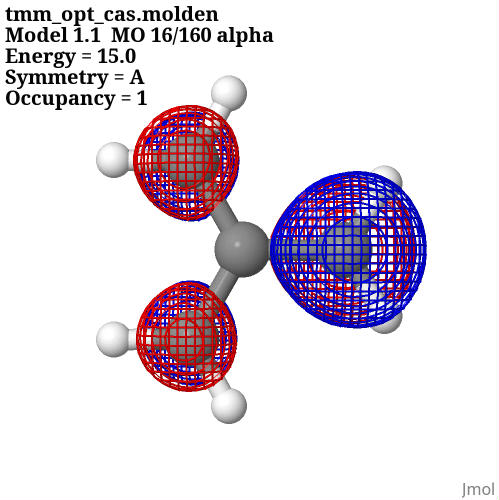} &
		\includegraphics[width=4cm]{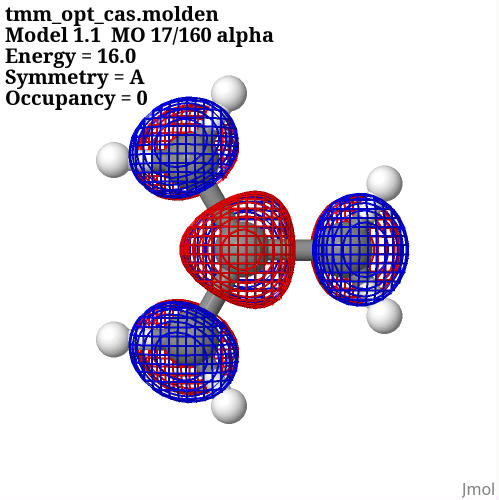} \\
	\end{tabular}
	\caption{trimethylenemethane (TMM).}
	\label{molecules}
\end{figure*}

\begin{figure*}[h]
	\begin{tabular}{cccc}
		\includegraphics[width=4cm]{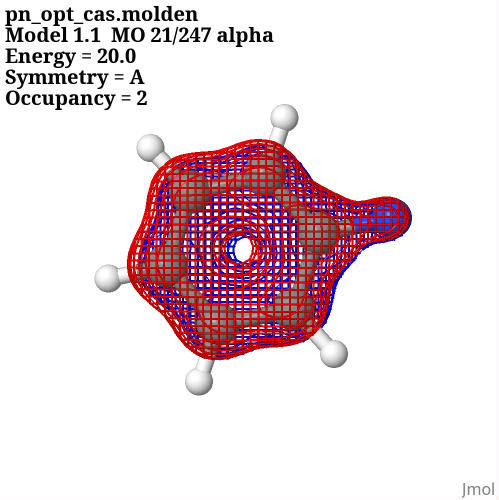} & \includegraphics[width=4cm]{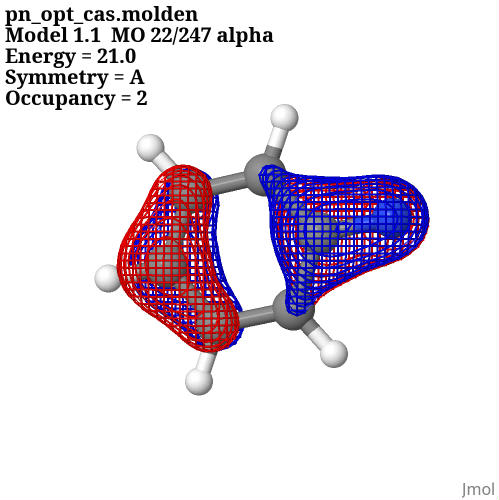} &
		\includegraphics[width=4cm]{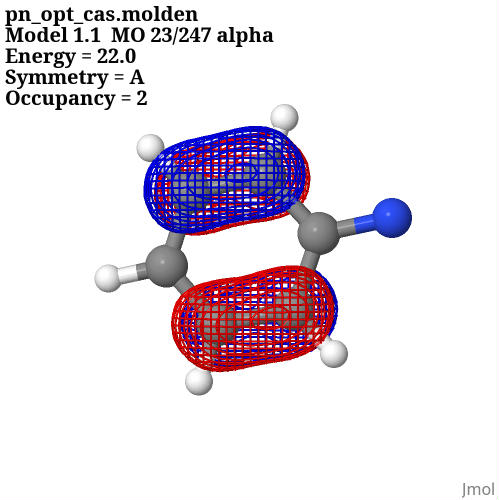} & \includegraphics[width=4cm]{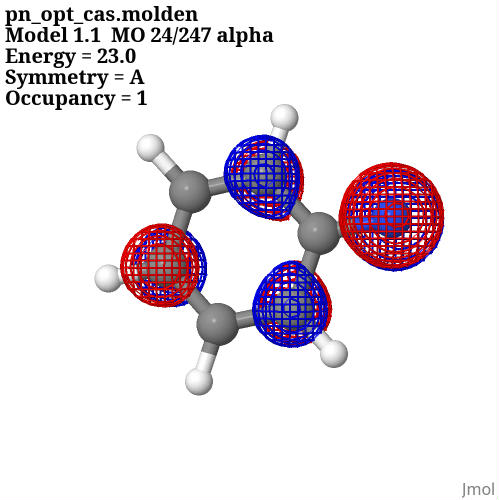} \\
		& & \\
		\includegraphics[width=4cm]{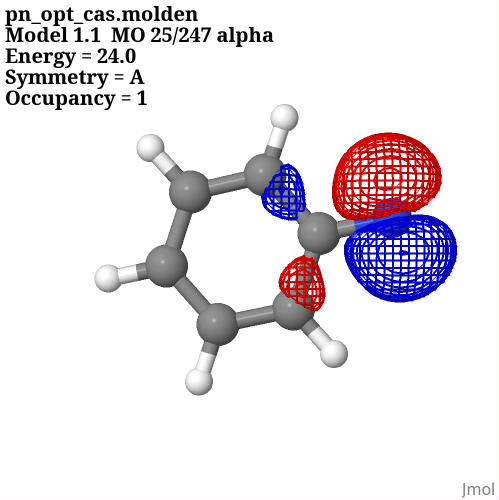} & \includegraphics[width=4cm]{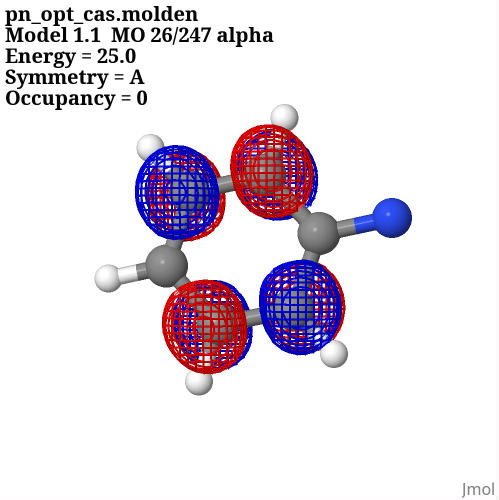} &
		\includegraphics[width=4cm]{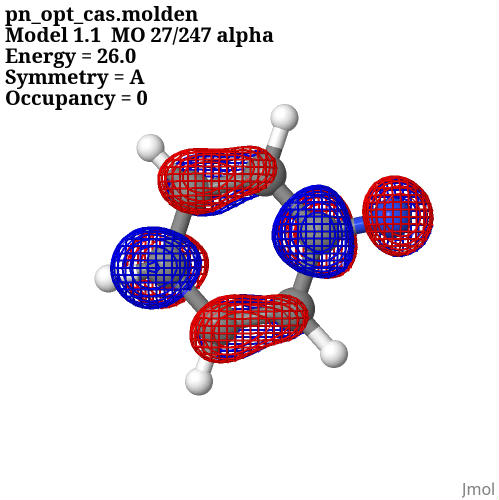} & \includegraphics[width=4cm]{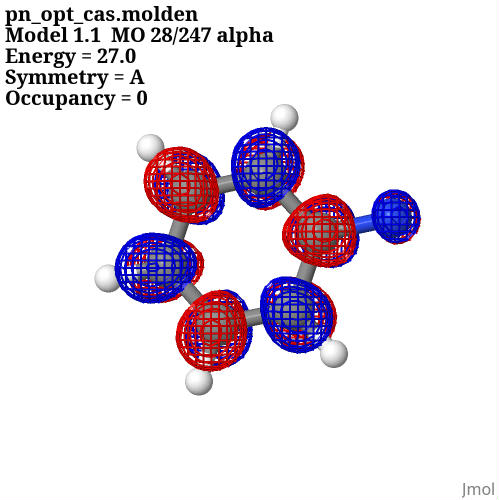} \\
	\end{tabular}
	\caption{phenylnitrene (PN)}
	\label{molecules}
\end{figure*}

\begin{figure*}[h]
	\begin{tabular}{cccc}
		\includegraphics[width=4cm]{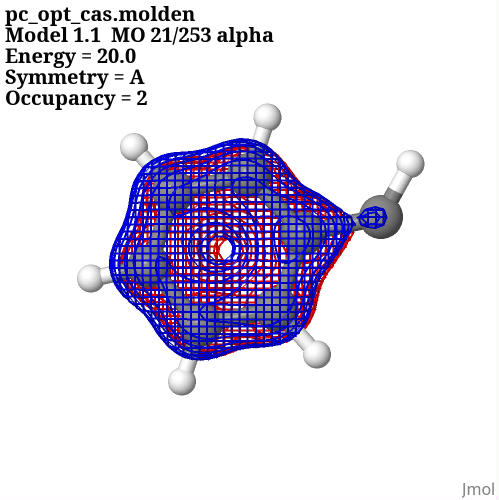} & \includegraphics[width=4cm]{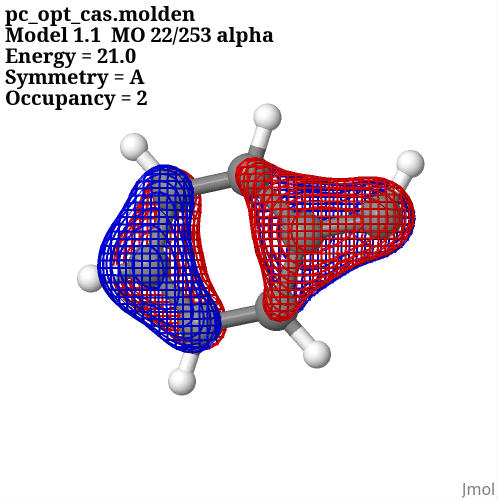} &
		\includegraphics[width=4cm]{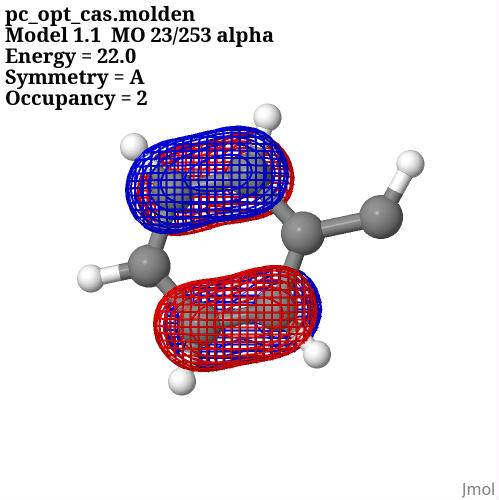} & \includegraphics[width=4cm]{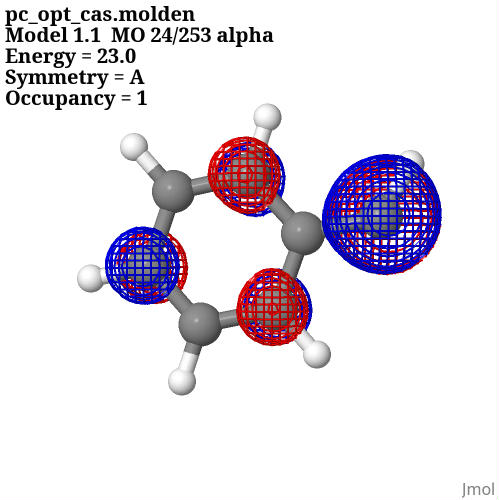} \\
		& & \\
		\includegraphics[width=4cm]{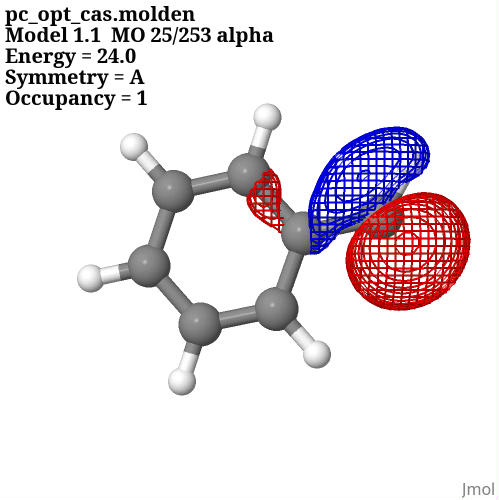} & \includegraphics[width=4cm]{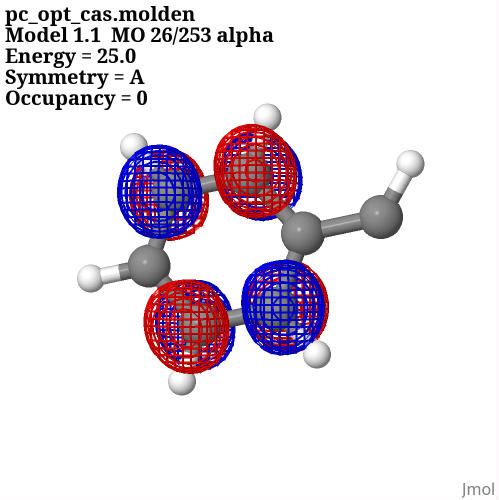} &
		\includegraphics[width=4cm]{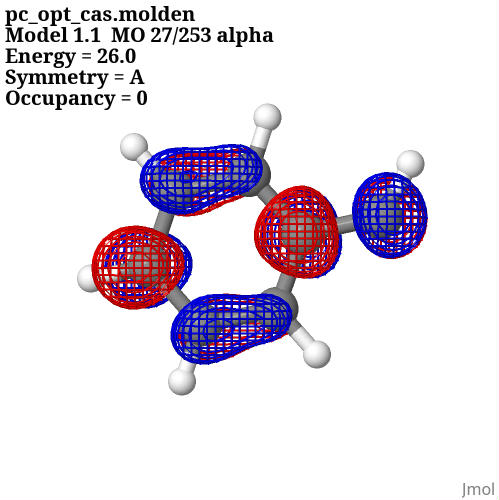} & \includegraphics[width=4cm]{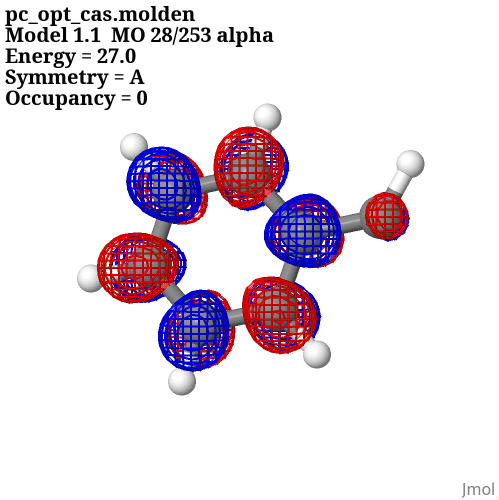} \\
	\end{tabular}
	\caption{phenylcarbene (PC).}
	\label{molecules}
\end{figure*}

\begin{figure*}[h]
	\begin{tabular}{cccc}
		\includegraphics[width=4cm]{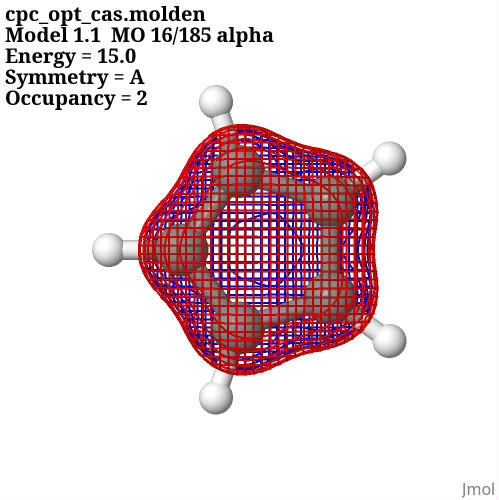} & \includegraphics[width=4cm]{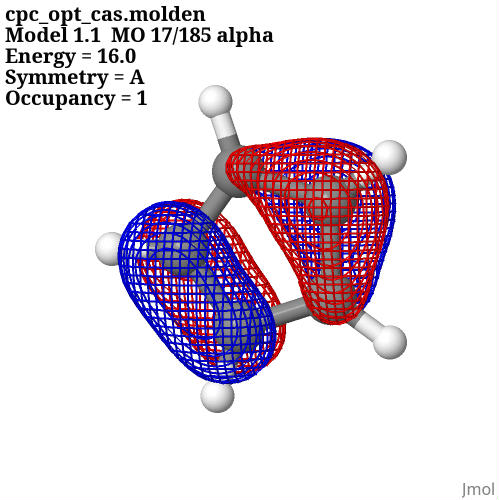} &
		\includegraphics[width=4cm]{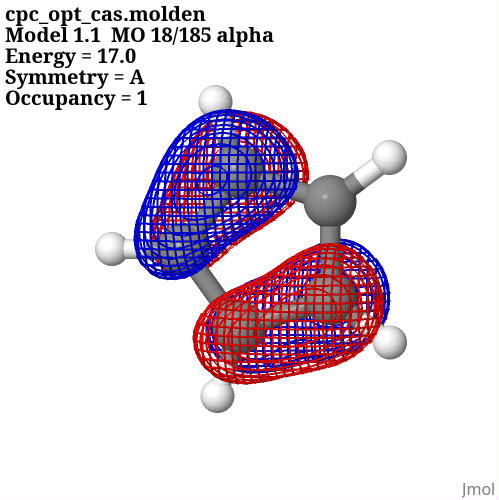} \\
		& & \\
		\includegraphics[width=4cm]{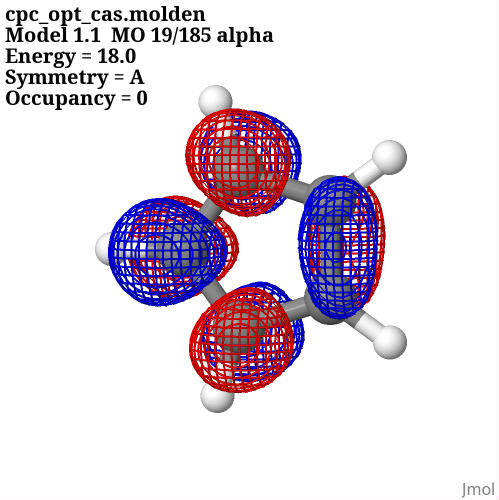} &  \includegraphics[width=4cm]{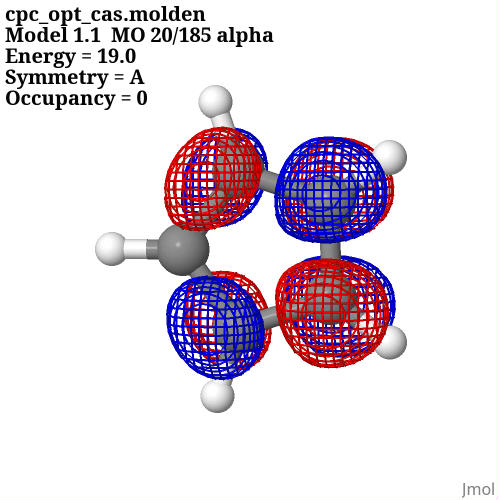} \\
	\end{tabular}
	\caption{cyclopentane cation (CPC).}
	\label{molecules}
\end{figure*}

\end{widetext}


\nocite{*}


\begin{thebibliography}{99}

\bibitem{Scully} 
G. Chen, S.A. Chin, Y. Dou, K.T. Kapale, M. Kim,
A.A. Svidzinsky, K. Urtekin, H. Xiong, and M. O. Scully,
Advances in Atomic, Molecular and Optical Physics {\bf 51}, 93 (2005).

\bibitem{Malrieu}
J.P. Malrieu, R. Caballol, C.J. Calzado, C. de Graaf, and N. Guih\'ery, 
Chem. Rev. {\bf 114}, 429 (2014).

\bibitem{Broer}
C. de Graaf and R. Broer, "{\it Magnetic Interactions in Molecules and Solids}", Springer International Publishing Switzerland (2016).

\bibitem{Abe}
M. Abe, Chem. Rev. {\bf 113}, 7011 (2013).

\bibitem{Stuyver}
Thijs Stuyver, Bo Chen, Tao Zeng, Paul Geerlings, Frank De Proft and Roald Hoﬀmann, Chem. Rev. {\bf 119}, 11291 (2019).

\bibitem{Loss}
G. Burkard, D. Loss and D.P. DiVincenzo, Phys. Rev. B {\bf 59}, 2070 (1999).

\bibitem{Burkard}
G. Burkard, T.D. Ladd, A. Pan, J.M. Nichol and J.R. Petta, Rev. Mod. Phys. {\bf 95}, 025003 (2023).

\bibitem{Hu}
X. Hu and S. Das Sarma, Phys. Rev. A {\bf 61}, 062301 (2000).

\bibitem{Calzado}
C.J. Calzado, J. Cabrero, J.P. Malrieu and R. Caballol, J. Chem. Phys. {\bf 116}, 3985 (2002).

\bibitem{Bohm1}
D. Bohm, D. Pines,  Phys. Rev. {\bf 82}, 625 (1951).

\bibitem{Bohm2}
D. Pines, D. Bohm, Phys. Rev. {\bf 85}, 338 (1952).

\bibitem{Bohm3}
D. Bohm, D. Pines, Phys. Rev. {\bf 92}, 609 (1953).

\bibitem{Ball}
A.D. McLachlan, M.A. Ball, Rev. Mod. Phys. {\bf 36}, 844 (1964).

\bibitem{Dunning}
T.H. Dunning and V. McKoy, J. Chem. Phys. {\bf 47}, 1735 (1967).

\bibitem{Furche1}
F. Furche, Phys. Rev. B {\bf 64}, 195120 (2001).

\bibitem{Furche2}
F. Furche, J. Chem. Phys. {\bf 129}, 114105 (2008).

\bibitem{Furche3}
H. Eschuis, J.E. Bates, and F. Furche, Theor. Chem. Ass., {\bf 131}, 1084 (2012).

\bibitem{Hesselmann}
A. He{\ss}elmann \& A. G\"orling, Molecular Physics {\bf 109}, 1 (2011).

\bibitem{Rinke}
X. Ren, P. Rinke, C. Joas, M. Scheffler, J. Mater. Sci. {\bf 47}, 7447 (2012).

\bibitem{Paris}
B. Helmich-Paris, J. Chem. Theory Comput. {\bf 15}, 4170 (2019).

\bibitem{Yang}
Y. Yang, D. Peng, E.R. Davidson, and W. Yang, J. Phys. Chem. A {\bf 119}, 4923 (2015).

\bibitem{Neese}
F. Neese, J. Phys. Chem. Sol.  {\bf 65},  781 (2004).

\bibitem{Yang_ocigoacenes}
Y. Yang, E. R. Davidson, and W. Yang, Proc. Natl. Acad. Sci. U.S.A {\bf 113}, E5098 (2016).

\bibitem{Borden}
 W.C. Lineberger  and  W.T. Borden, Phys. Chem. Chem. Phys. {\bf 13}, 11792 (2011).  

\bibitem{Dressler}
J.J. Dressler,  A. C\'ardenas Valdivia,  R. Kishi,  G.E. Rudebusch,  A.M. Ventura,
B.E. Chastain,  C.J. G\'omez-Garci\'a,  L.N. Zakharov,  M. Nakano,
J. Casado,  and M.M. Haley, Chem {\bf 6}, 1353 (2020).

\bibitem{Petta}
Petta, J. R., A. C. Johnson, J. M. Taylor, E. Laird, A. Yacoby, M. D.
Lukin, C. M. Marcus, M. P. Hanson, and A. C. Gossard, 
Science {\bf 309}, 2180 (2005).

\bibitem{Krylov}
L.V. Slipchenko and A.I. Krylov, J. Chem. Phys. {\bf 117}, 4694 (2002).

\bibitem{Krylov_TMM}
L.V. Slipchenko and A.I. Krylov, J. Chem. Phys. {\bf 118}, 6874 (2003).

\bibitem{Lee}
J. Lee and M. Head-Gordon, J. Chem. Phys. {\bf 150}, 244106 (2019).

\bibitem{Wang}
F. Wang, T. Ziegler,  J. Chem. Phys. {\bf 122}, 074109 (2005).

\bibitem{Nakano}
M. Nakano and B. Champagne, J. Chem. Phys. {\bf 138}, 244306 (2013).

\bibitem{He}
L. He, G. Bester, and A. Zunger, Phys. Rev. B {\bf 72}, 195307 (2005).

\bibitem{Hay}
P.J. Hay, J.C. Thibeault, R.J. Hoffman, J. Am. Chem. Soc. {\bf 97}, 4884 (1975).

\bibitem{Sun}
Q. Sun, J. Yang, G. Kin-Lic Chan, Chem. Phys. Lett. {\bf 683}, 291 (2017).

\bibitem{Sun1}
Q. Sun {\it et al}, J. Chem. Phys. {\bf 153}, 024109 (2020).

\bibitem{Sun2}
Q. Sun {\it et al}, WIREs Comput. Mol. Sci.  {\bf 8}, e1340 (2018).

\bibitem{Sun3}
Q. Sun, J. Comp. Chem. {\bf 36}, 1664 (2015).

\bibitem{Wang2}
L.-P. Wang, C. Song, J. Chem. Phys. {\bf 144}, 214108 (2016). 

\bibitem{Schafer}
A. Sch\"afer, C. Huber, R. Ahlrichs, J. Chem. Phys. {\bf 100}, 5829–5835 (1994).

\bibitem{Li}
P.J. Lestrange, F. Egidi, X. Li, J. Chem. Phys. {\bf 143}, 234103 (2015).

\bibitem{Sheng}
H. Sheng, X. Ma, H.-R. Lei, J. Milton, W. Tang,  C. Jin, J. Gao, A.M. Wittrig, E.F. Archibold,  J.J. Nash, and H.I. Kentt\"amaa,
ChemPhysChem {\bf 19}, 2839 (2018).

\bibitem{W}
P.G. Wenthold, R.R. Squires, and W.C. Lineberger, J. Am. Chem. Soc. {\bf 120}, 5279 (1998).

\bibitem{Lindhard}
J. Lindhard  {\it "On the properties of a gas of charged particles"}, Danske Matematisk-fysiske Meddelelser {\bf 28} 1, (1954). 

\bibitem{Pulay}
P. Pulay, T.P. Hamilton, J. Chem. Phys. {\bf 88}, 4926 (1988).


\end{thebibliography}
\end{document}